\documentclass[prd,aps,showpacs]{revtex4}
\usepackage{amsmath}
\usepackage{graphicx}
	\usepackage{amssymb}

\begin{document}
 
\title{Analytic approximation to 5 dimensional Black Holes with one compact dimension}
\author{D. Karasik}
\author{C. Sahabandu}
\author{P. Suranyi}
\author{L. C. R. Wijewardhana}
\affiliation{Department  of Physics,
University of Cincinnati, Cincinnati, Ohio}

\begin{abstract} We study black hole solutions in $R^4\times S^1$ space, using an expansion to fourth order in the ratio of the radius of the horizon, $\mu$, and the circumference of the compact dimension, $L$. A study of geometric and thermodynamic properties indicates that the black hole fills the space in the compact dimension at $\epsilon(\mu/L)^2\simeq0.1$. At the same value of $\epsilon$ the entropies of the uniform black string and of the black hole are approximately equal. 
\end{abstract}
\pacs{04.50.+h, 04.70.Bw, 04.70.Dy}
\maketitle

\section{Introduction}
If the topology of space-time  is $R^{4}\times S^{1}$ then the only known exact solution representing a black object is the uniform black string with horizon topology
$S^{2}\times S^{1}$~\cite{blackstring}.  Though this solution exists for all values of the mass it is unstable below a critical value, $M_{\rm GL}$,
as shown by Gregory and Laflamme \cite{GL}.

Horowitz and Maeda \cite{horowitz} argued that a uniform black string cannot change its
topology into a black hole in finite affine time, making the possibility of such a transition questionable.  They suggested the possibility of a transition to a nonuniform black string. Gubser \cite{gubser} showed the existence of  nonuniform black string solutions. The  non-uniform black string solution in 6 dimensions  was investigated 
numerically \cite{nonuniform} for a range of the mass values above $M_{\rm GL}$. 

Nonuniform black string configurations do not exist for masses below the Gregory-Laflamme point, in the region where the uniform black string solution is unstable~\cite{nonuniform}.  A natural candidate for a black object in this mass range is a black hole. Unfortunately, no exact black hole solutions are known in a 5 (or more) dimensional space with a compactified dimension.  Still, on the basis of physical intuition, a black hole solution should exist for very small values of the mass. When the radius of the horizon is much smaller than the size of the compactified dimension, i.e. when $M\to0$,  the black hole should be unaware of the compactification. Then a Myers-Perry~\cite{myers} solution should become asymtotically exact. Indeed, a numerical solution, extending the Myers-Perry solution to larger values of the mass was found by Harmark and Obers~\cite{linearized,harmark} and Kudoh and Wiseman~\cite{kudoh2}. 

Using general arguments Kol~\cite{kol2} suggested that the black hole branch and the nonuniform black string branch meet at a point when the black hole fills the compact dimension. Further arguments for such a transition were presented in~\cite{wiseman2,kol}.

Recently, we have studied a related problem, namely the existence of black holes in  Randall-Sundrum~\cite{rs1,rs2} theories.
We used~\cite{us1,us2} an approximation scheme based on the expansion of solutions in the ratio of the radius of the horizon of the black hole to the ADS curvature.  In this paper we employ a similar strategy by expanding the metric and other relevant quantities in the ratio of the two natural lengths associated with a black hole configuration, (i) the five dimensional Schwarzschild radius, $\mu$, associated
with the mass $\mu=\sqrt{8G_{5}M/(3\pi)}$ and (ii) the
compactification length, $L$, defined as the proper circumference
of the compact dimension in the region far away from the mass. The dimensionless ratio of these two quantities serves as an excellent expansion parameter. As the solution is an even function of $\mu$ and $L$ we use 
$
\epsilon=\left(\frac{\mu}{L}\right)^2 
$ as our expansion parameter. Such an expansion for ADD black holes has recently been proposed and Harmark~\cite{harmark2} and Gorbonos and Kol~\cite{gorbonos} evaluated the leading term of the expansion. 

To find an unique solution we must fix boundary conditions, both at infinity and at the horizon of the black hole.  We find  solutions in two different regions.
In the asymptotic region, $r\gg\mu$ an `asymptotic solution',
and in the near horizon region, $r\ll L$ a `near solution' is found.
Using the asymptotic solution we 
satisfy  asymptotic boundary conditions but the
boundary conditions at the horizon cannot be satisfied. The near solution suffers from the opposite
problem, as it cannot be used to satisfy asymptotic boundary conditions.
To solve those problems we combine the two solutions. We match the
asymptotic solution with the near solution in the region $\mu\ll r\ll L$
where both are valid.
This method was used in \cite{us1,us2} to find small black holes in Randall-Sundrum
scenario, and in \cite{kol} to find small black holes on cylinders. The work
in this paper is parallel to and agrees with previous results~ \cite{harmark2,gorbonos}, but here we calculate the metric up to
fourth order in $\mu/L$ (second order in $\epsilon$). 
The organization of this paper is as follows:  In section \ref{sec:method} we describe
the general method of our calculations, and list the parameters that will be calculated
from the metric. In sections \ref{sec:zeroth}-\ref{sec:second}
we present the detailed calculations up to second order.  In section \ref{sec:summary} we summarize our results for the metric,
up to second order, and discuss the possible scenarios for  black holes of increasing mass. A reader not interested in technical details should read the next section, (\ref{sec:method}) and then proceed directly to the summary and discussions (\ref{sec:summary}).

\section{The general method}
\label{sec:method} As we indicated in the Introduction we will investigate black hole solutions of the Einstein equation in 5 dimensional space when one of the dimensions is compactified.  Recently, this problem has attracted the attention of several groups.  Harmark and Obers\cite{harmark,kol} introduced the relative tension of black holes as an order parameter, wrote down a generalized Smarr formula~\cite{smarr}, investigated the phase diagram for black objects, and studied the solutions numerically.  Gorbonos and Kol~\cite{gorbonos} proposed an analytic approximation scheme based on the expansion in the ratio of the radius of the horizon and the compactification length. This method is similar to the expansion method we used~\cite{us1,us2} to investigate black holes in the Randall-Sundrum \cite{rs1,rs2} scenario.  We will follow a similar path here and calculate further terms of the expansion.

By necessity, the perturbation method is applied in two overlapping
regions. In the asymptotic region, $r\gg\mu$, $\mu$ is the
smallest scale. Therefore, an expansion in $\mu/L$ is actually an
expansion in $\mu$ and in every  order the metric is a general function of the
dimensionless coordinates $x/L$,
\begin{equation}
g_{\alpha\beta}(x)=\sum_{n=0}(\mu/L)^{n}g^{Asymp}_{\alpha\beta,n}(x/L).
\label{asym}
\end{equation}
We refer to this
solution as the `asymptotic solution.'  

In the near horizon region,
$r\ll L$, $L^{-1}$ is the smallest scale. Therefore, an expansion
in $\mu/L$ is actually an expansion in $L^{-1}$. In every order, the metric is a general function of  the dimensionless coordinates $x/\mu$,
\begin{equation}
g_{\alpha\beta}(x)=\sum_{n=0}(\mu/L)^{n}g^{Near}_{\alpha\beta,n}(x/\mu).
\label{near}
\end{equation}
We refer to this
solution as the `near solution.'

\subsection{The equations}
 In each region, when we calculate
the $n$-th order terms, Einstein's equations is solved in terms of two
functions: A gauge function and a wave function, which satisfies a
linear, (inhomogeneous), master equation. The differential
operator for the master equation in the asymptotic region is
\begin{equation}
    \frac{\partial^{2}}{\partial r^{2}}+\frac{2}{r}\frac{\partial}{\partial r}
    +\frac{\partial^{2}}{\partial w^{2}}~,
    \label{masterasymp}
\end{equation}
where $w$ is a coordinate in the compact dimension and $r$ is a radial
coordinate in 3-dimensional space.
In the near region the operator is
\begin{equation}
    (R^{2}-1)\left(\frac{\partial^{2}}{\partial R^{2}}-\frac{1}{R}\frac{\partial}{\partial R}
    \right)+\frac{\partial^{2}}{\partial \psi^{2}}
    -4\cot2\psi\frac{\partial}{\partial \psi}-3~,
    \label{masternear}
\end{equation}
where $R$ is a radial coordinate and $\psi$ is an angular coordinate in 4-dimensional space. The driving term in the master equation for the $n$th order contribution depends on the solution in lower orders.
It determines an inhomogeneous solution in the wave function. The additional homogeneous solution
introduces new parameters that should be fixed by boundary conditions.

\subsection{The boundary conditions.}
In empty space, or for the uniform black string configuration, the cylinder is invariant under translations in the
compact direction.
When a point mass is introduced translation invariance is broken, but
 a $Z_{2}$ reflection symmetry remains around the location of the mass in the compact direction.
So, if we endow the compact direction with the coordinate $w\in[-L/2,L/2]$
and put the mass at $w=0$ then the solution is symmetric about $w\rightarrow-w$
and periodic in $w$ with period $L$.

Asymptotically (far away from the mass), we assume that the metric is Minkowski.
The leading order correction to the Minkowski metric is of order $r^{-1}$. Such a term provides, among others,
 the four-dimensional Newtonian potential.

We must impose constraints on the near solution, as well.  The horizon of a small black hole has $S^{3}$ topology.
We require that there is no black string attached to the horizon.
Consequently, the Kretchmann scalar must be regular on the $w$ axis everywhere, when $w\not=0$. Furthermore,  the surface gravity must be constant, otherwise the horizon is not regular~\cite{wald}.

\subsection{Matching the asymptotic and near solutions}
To match the asymptotic and the near
solutions we must work in the intermediate region $\mu\ll\rho\ll L$,
where $\rho$ is a four-space-dimensional radial coordinate. In this
region the functions $g_{\alpha\beta,n}^{\rm Asymp}(x/L)$ can be expanded in the coordinates. Owing to the $S^2\times S^1$ symmetry the components  depend only on $\rho$ and  $\psi$, the angle between the compact direction and the
four-dimensional sub-manifold.   Then one can use a double expansion in $\mu/\rho$ and $\rho/L$ to write the metric as
\begin{equation}
    g_{\alpha\beta}(\rho,\psi)=\sum_{n=0}^{\infty}\sum_{k=0}^{\infty}
    g_{\alpha\beta}^{(n,k)}(\psi)\left(\frac{\mu}{\rho}\right)^{2n}
    \left(\frac{\rho}{L}\right)^{2k}~,
    \label{double}
\end{equation}
The expansion includes only even
powers due to the $Z_{2}$-symmetry as will be explained later.

Globally, our expansion parameter is $\epsilon=(\mu/L)^{2}$.
In the two regions $\epsilon$ has a different meaning. 
The asymptotic
solution is constructed as an expansion in $\mu$, such that
$(\mu/\rho)^{2n}(\rho/L)^{2k}=\epsilon^{n}(\rho/L)^{2(k-n)}$, so
one keeps $n$ constant and solves the equations for all values of $k$. For example, Minkowski metric corresponds to the zeroth order ($n=0$,
 $k=0\ldots\infty$) approximation to our problem, with appropriate coefficients, $ g_{\alpha\beta}^{(0,k)}$. The first order, $n=1$ ,$k=0\ldots\infty$, provides linearized gravity (Newtonian potential), etc.  

\begin{figure}[htbp!]
    \centering
    \includegraphics[scale=1]{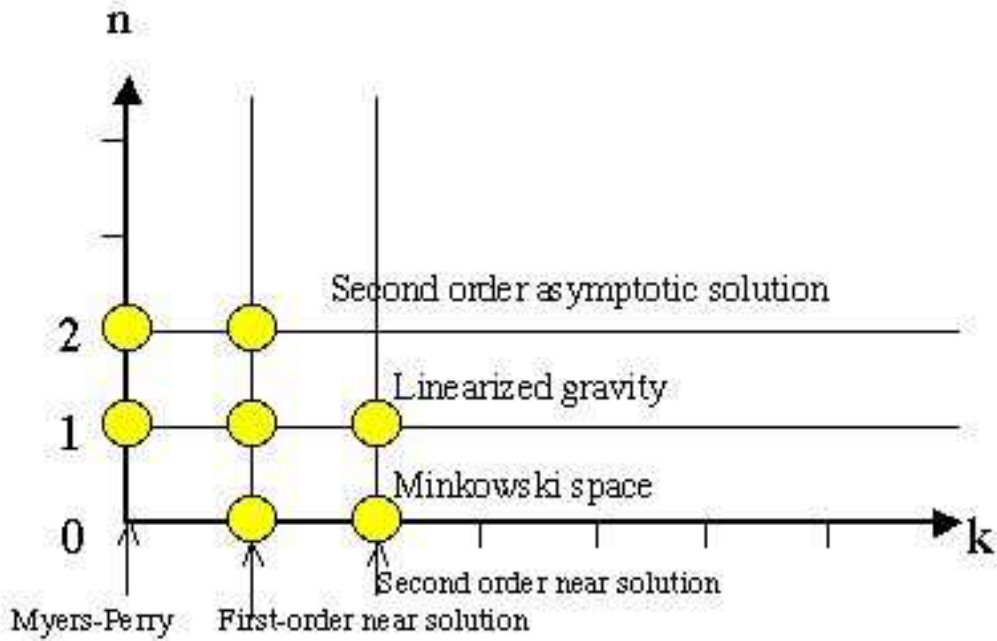}
    \caption{
    $(\mu/\rho)^{2n}(\rho/L)^{2k}$ terms as points in the $(n,k)$
    grid. The horizontal (vertical) lines describe the asymptotic
    (near) solution in increasing order of $\epsilon$.
    The matching is done at the intersecting points denoted by circles.}
\label{gridfig}
\end{figure}

The
near solution is constructed as an expansion in $L^{-1}$, such
that
$(\mu/\rho)^{2n}(\rho/L)^{2k}=\epsilon^{k}(\mu/\rho)^{2(n-k)}$, so
at fixed $k$ we solve for all values of $n$. Here, $k=0$, with appropriately chosen coefficients,  $ g_{\alpha\beta}^{(0,k)}$, corresponds to the Myers and Perry five dimensional black hole (MPS)
\cite{myers}. The matching is done in the intersecting points denoted by circles, as
can be seen in Fig.(\ref{gridfig}).
For example, first order of
the asymptotic solution intersects zeroth order of the near
solution at $(n=1,k=0)$. We use this information to fix
parameters in the asymptotic solution.
\begin{figure}[htb]\centering
    \includegraphics[scale=1]{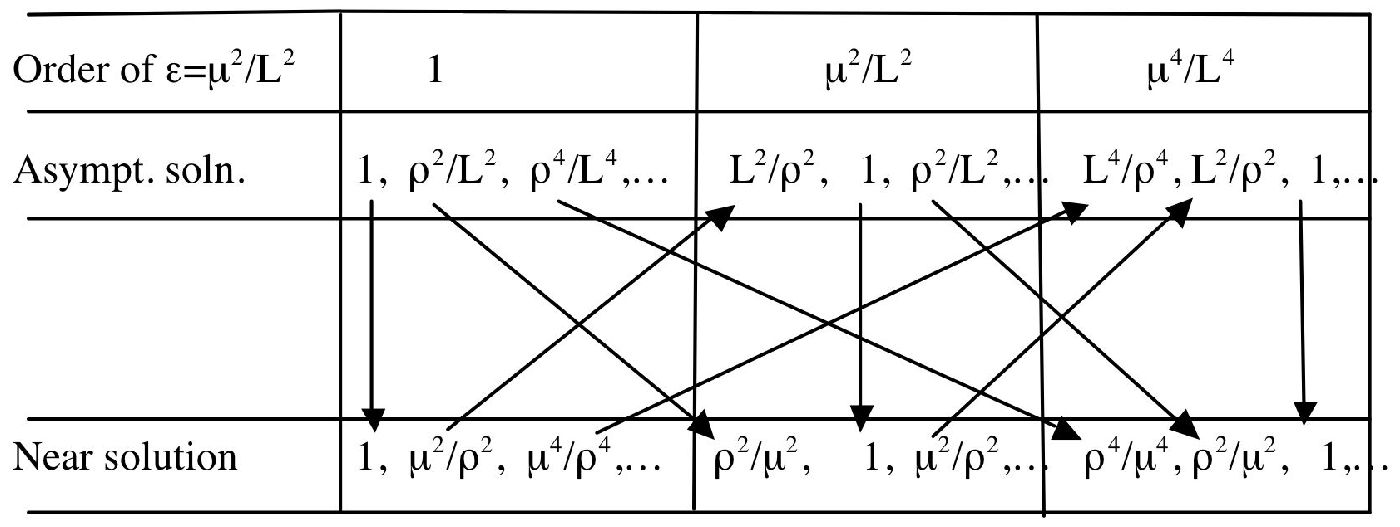}
    \caption{\label{tablefig}In the intermediate region the asymptotic solution is expanded in $\rho/L$, while the
    near solution is expanded in $\mu/\rho$. The table shows the sequence of calculating terms
  in the asymptotic and near solutions. Terms connected by an arrow are identified by the double expansion. A similar table has been presented in \cite{gorbonos}}
\end{figure}
The calculation is carried out order by order. Both solutions (asymptotic and near)
must be known up to order $\epsilon^{n-1}$ before calculating order $\epsilon^{n}$.
Within order $\epsilon^{n}$, one should calculate the asymptotic solution first and then
the near solution. The sequence of  calculation of terms in the two solutions is depicted in Fig.(\ref{tablefig}).
\subsection{Physical parameters}
Two parameters  can be measured in the asymptotic region,
the ADM mass and the relative tension around the compact dimension.~\cite{harmark}
Expanding the asymptotic metric around the Minkowski metric we write
$g_{\mu\nu}=\eta_{\mu\nu}+h_{\mu\nu}$,
Then the ADM mass, $M$, and the tension, $b$, are defined as
\begin{eqnarray}
        M&=&\frac{1}{4G_{5}}\lim_{r\rightarrow\infty}r^{2}
        \int_{-L/2}^{L/2}\left[\frac{2}{r}h_{rr}-h_{ww,r}\right]dw~,
    \label{defADMmass}\\
    b&=&\int d^{4}x T^{ww}=-\frac{1}{4G_{5}}\lim_{r\rightarrow\infty}r^{2}
        \int_{-L/2}^{L/2}\left[\frac{2}{r}h_{rr}+h_{tt,r}\right]dw~.
    \label{deftension}
\end{eqnarray}
It is useful to define the relative tension, $n=-b/M$.
In \cite{harmark} the pair $(M,n)$ is used to describe a phase diagram of all
possible black objects.

The near solution provides the thermodynamic parameters of the horizon, i.e.
entropy and temperature. The entropy is defined as the three dimensional 'area' of
the horizon divided by $4G_{5}$. The temperature is defined through the zeroth
law of black holes thermodynamics as the surface gravity divided by $2\pi$.
A Smarr formula \cite{harmark,kol} relates the entropy, temperature, mass, and relative tension as~\cite{harmark}
\begin{equation}
    TS=\frac{2-n(M)}{3}M~.\label{Smarrint}
\end{equation}

\section{Zeroth order in $\epsilon$}
\label{sec:zeroth} In the asymptotic region zeroth order in
$\epsilon$ means that we neglect the mass. Therefore, the
configuration is asymptotically flat (since the mass is confined
to a small region). So, the asymptotic form of the metric is
Minkowski on a cylinder
\begin{equation}
    ds^{2}=-dt^{2}+dr^{2}+dw^{2}+r^{2}(d\theta^{2}+\sin^{2}\theta d\phi^{2})~,
    \label{A0}
\end{equation}
where $w$ is the compact coordinate such that $w\in[-L/2,L/2]$.

In the near region zeroth order in $\epsilon$ means that $L\to\infty$.  The mass point is in five dimensional
asymptotic flat infinite manifold. Therefore, the near solution is given by the
Myers-Perry~\cite{myers} solution (MPS),
\begin{equation}
    ds^{2}=-(1-\frac{\mu^{2}}{\rho^{2}})dt^{2}
    +\frac{d\rho^{2}}{1-\frac{\mu^{2}}{\rho^{2}}}+\rho^{2}d\psi^{2}
    +\rho^{2}\sin^{2}\psi(d\theta^{2}+\sin^{2}\theta d\phi^{2})~,
    \label{N0}
\end{equation}
where $\mu$ is the five dimensional Schwarzschild radius. In order
to match the metrics (\ref{A0}, \ref{N0}) one needs to specify a
coordinate transformation between $(t,r,w,\theta,\phi)$ and
$(t,\rho,\psi,\theta,\phi)$. We choose to work with a specific
coordinate transformation throughout the calculation of all
orders. (One can, however, choose to redefine the transformation
in each order). The coordinates $t$, $\theta$, and
$\phi$ are shared by the coordinate systems. Furthermore, we want to preserve the periodic structure in
$w$ so we choose the transformation
\begin{subequations}
\label{transformation}
\begin{eqnarray}
    v&=&\frac{L}{\pi}\sin\frac{\pi w}{L}=\rho\cos\psi~,\;\;w\in[-L/\pi,L/\pi]~,
    \label{z}\\
    r&=&\rho\sin\psi~,\label{r}\\
    \rho&=&\sqrt{r^{2}+v^{2}}=\sqrt{r^{2}+\frac{L^{2}}{\pi^{2}}\sin^{2}\frac{\pi w}{L}}
    ~,\label{rho}\\
    \psi&=&\arctan\frac{r}{v}=\arctan\frac{\pi r}{L\sin(\pi w/L)}~,\label{psi}
\end{eqnarray}
\end{subequations}
One can verify that the metrics (\ref{A0},\ref{N0}) agree up to
order $\mu^{0}(1/L)^{0}$.

We  show now that the expansion contains only terms with integer powers of $(\mu/L)^{2}\equiv\epsilon$. It is sufficient to establish that the solution must be an even function of $\mu$ and of $L$ (at fixed $\mu^2\sim MG_5$).  The solution is an even function of $\mu$, because  the driving term, the zeroth order near metric (\ref{N0}) depends only on $\mu^{2}$. The zeroth order near metric, (\ref{N0}), combined with (\ref{transformation}) is also an even function of $L$.   The conditions imposed on the metric  also force it to be even in $L$. The mass is located at $\rho=0$, ($r=w=0$). The configuration of a
mass point on a cylinder has a $Z_{2}$ symmetry about $w=0$,
($\psi=\pi/2$),  requiring that the terms of the metric depend on $w$ and $L$ through functions of the form $\cos(2\pi n w/L)$.

\section{First order in $\epsilon$}
\label{sec:first}
\subsection{The asymptotic metric}
In the asymptotic region, $r\gg\mu$, we expand the metric as
\begin{equation}
    g_{\alpha\beta}=\eta_{\alpha\beta}+\epsilon h_{\alpha\beta}~,
    \label{A1}
\end{equation}
where $\eta_{\alpha\beta}$ is the flat metric (\ref{A0}).

 The general form of the
solution for $h_{\alpha\beta}$ appears in many places in the literature
\cite{linearized}.  We re-derive it here using the
matching method. We expand the Einstein equation to first order in
$\epsilon$ and get a set of homogeneous linear equations for
$h_{\alpha\beta}$. Using the coordinate transformation
$w\rightarrow w+\epsilon W(r,w)$ we choose the gauge
$h_{rw}=W_{,r}(r,w)$. Next we solve $G_{rw}=G_{ww}=0$ for
$h_{rr}$, and $G_{rr}=0$ for $h_{ww}$, where $G_{\mu\nu}$ is the Einstein tensor. The solution is given by
\begin{subequations}
\label{h1}
\begin{eqnarray}
    h_{rr}&=&\frac{\alpha}{r}-\frac{r}{2}h_{tt,r}~,\\
    h_{ww}&=&-\frac{\alpha}{r}-\frac{r}{2}h_{tt,r}+2W_{,w}(r,w)~,\\
    h_{rw}&=&W_{,r}(r,w)~,\\
    h_{\theta\theta}&=&h_{\phi\phi}=0~,
\end{eqnarray}
\end{subequations}
where the gauge function satisfies the symmetry properties $W(r,w+L)=W(r,w)=-W(r,-w)$ and $\alpha$ is a constant.
The function $h_{tt}$ satisfies the equations
\begin{eqnarray}
    & &h_{tt,rr}+\frac{2}{r}h_{tt,r}+h_{tt,ww}=0~,\nonumber\\
    & &h_{tt}(r,w)=h_{tt}(r,-w)\;\;;\;h_{tt}(r,w+L)=h_{tt}(r,w)~.
    \label{httEq}
\end{eqnarray}
The solution to Eq.(\ref{httEq}) can be written as a Fourier series
\begin{equation}
    h_{tt}(r,w)=\sum_{n=0}^{\infty}c_{n}\frac{\exp(-2\pi n r/L)}{r}
    \cos(2\pi n w/L)~.
    \label{httmodes}
\end{equation}
The constants $c_{n}$ are determined by matching (\ref{httmodes}) to the zeroth order near solution (\ref{N0}).
Then it follows that up to leading order in $L^{-1}$ the two solutions
must be identical, $\epsilon h_{tt}=\mu^{2}\rho^{-2}$.
The constants $c_{n}$ are given by the Fourier expansion of $\mu^{2}\rho^{-2}$.
\begin{subequations}
\label{Fourier1}
\begin{eqnarray}
        \epsilon \frac{c_{0}}{r}=\frac{1}{L}\int_{-L/2}^{L/2}
    \frac{\mu^{2}}{r^{2}+(L/\pi)^{2}\sin^{2}(\pi w/L)}dw
    &\Longrightarrow& c_{0}=\pi L +{\cal O}(L^{0})~,\\
    \epsilon \frac{c_{n}}{r}\exp(-2\pi n r/L)
    =\frac{2}{L}\int_{-L/2}^{L/2}\frac{\mu^{2}}{r^{2}+(L/\pi)^{2}\sin^{2}(\pi w/L)}
    \cos(2\pi n w/L)dw
    &\Longrightarrow& c_{n}=2\pi L +{\cal O}(L^{0})~.
\end{eqnarray}
\end{subequations}
Using (\ref{Fourier1}) we are able to
sum Eq.(\ref{httmodes}) as
\begin{equation}
    h_{tt}(r,w)=\frac{\pi L}{r}\frac{\sinh\frac{2\pi r}{L}}
    {\cosh\frac{2\pi r}{L}-\cos\frac{2\pi w}{L}}~.
    \label{htt1final}
\end{equation}

For $r\gg L$ the metric is given in terms of the `zero mode,'  independent of $w$, and can be summarized as
\begin{subequations}
\label{h1asymp}
\begin{eqnarray}
    g_{tt}&\sim&-1+\epsilon\frac{\pi L}{r}~,\\
    g_{rr}&=&1+\epsilon\frac{2\alpha-\pi L}{2r}~,\\
    g_{ww}&=&1+\epsilon\left[-\frac{2\alpha+\pi L}{2r}+2W_{,w}(r,w)\right]~,\\
    g_{rw}&=&\epsilon W_{,r}(r,w)~.
\end{eqnarray}
\end{subequations}
The constant $\alpha$, which appears in the metric (\ref{h1asymp}) is related to the
tension in the compact dimension
\begin{equation}
    b=\int d^{4}x T^{ww}=-\frac{\epsilon}{4G_{5}}\lim_{r\rightarrow\infty}r^{2}
        \int_{-L/2}^{L/2}\left[\frac{2}{r}h_{rr}+h_{tt,r}\right]dw
        =-\frac{\alpha\epsilon L}{2G_{5}}~.
    \label{tension}
\end{equation}
The tension $b$ should be zero because we deal here with
linearized gravity, with no interaction between the mass and its
periodic images.  Consequently, we must set $\alpha=0$. 

Dimensional analysis also requires $\alpha=0$. First, we can deduce from Eqs.(\ref{h1}) that $\alpha\propto L$. The asymptotic solution is an expansion in the mass, so $\alpha$ must be independent of the mass and proportional to $L$, the only quantity of the correct dimension at our disposal. As we have discussed earlier, function (\ref{htt1final})
is even in $L$. If $\alpha=c L\neq0$ then the functions $h_{rr}$ and $h_{ww}$, in (\ref{h1}), would have a mixed symmetry under $L\to-L$. Note that this symmetry argument can and will be used in higher ($n$th) orders  of the expansion to eliminate terms of the form $\alpha_n/r$ from $g_{rr}$ and $g_{ww}$.

Next, we turn to evaluate the conserved ADM mass using
Eqs.(\ref{h1asymp})
\begin{equation}
    M=\frac{\epsilon}{4G_{5}}\lim_{r\rightarrow\infty}r^{2}
        \int_{-L/2}^{L/2}\left[\frac{2}{r}h_{rr}-h_{ww,r}\right]dw
        =\frac{3\pi\mu^{2}}{8G_{5}}~.
    \label{ADMmass}
\end{equation}
The effective four-dimensional Newton constant is defined by
$g_{tt}=-1+2G_{4}M/r$. Using Eqs.(\ref{h1asymp}, \ref{ADMmass}) we
find that $G_{4}=4G_{5}/(3L)$. We should emphasize that the
physically measurable quantities, the relative tension, the mass, and
Newton's constant, are independent of the gauge function $W(r,w)$.

As we mentioned earlier, in higher orders of the expansion the
relative tension in the compact dimension is not zero. However, we
require that the ADM mass is completely determined by the first order,
such that higher orders do not change Eq.(\ref{ADMmass}).
This way, we make sure that the expansion in $\epsilon$ is also
an expansion in $M$ and that thermodynamic properties of
the black hole are well defined in each order. We will return
to this issue in section \ref{sec:second} when we discuss the second order contributions.

The gauge function $W(r,w)$, which appears in (\ref{h1}), can be partially determined
by matching to the zeroth order of the near solution. However, we
prefer to determine the function completely by matching to the first
order of the near solution, as well.

\subsection{The near solution}
In the near region it is convenient to use the coordinates
$(t,\rho,\psi,\theta,\phi)$ just like in Eq.(\ref{N0}).
We use  the following ansatz for
the metric:
\begin{equation}
    ds^{2}=-B(\rho,\psi)dt^{2}+\frac{A(\rho,\psi)}{B(\rho,\psi)}d\rho^{2}
        +2V(\rho,\psi)d\rho\,d\psi+\rho^{2}U(\rho,\psi)d\psi^{2}
        +\rho^{2}\sin^{2}\psi(d\theta^{2}+\sin^{2}\theta\,d\phi^{2})~.
    \label{spherical}
\end{equation}
The functions
$B,A,U,V$ are expanded to first order in $\epsilon$ as
\begin{subequations}
\label{N1}
\begin{eqnarray}
    B(\rho,\psi)&=&1-\frac{\mu^{2}}{\rho^{2}}-\epsilon B_{1}(\rho,\psi)~,\\
    A(\rho,\psi)&=&1+\epsilon A_{1}(\rho,\psi)~,\\
    U(\rho,\psi)&=&1+\epsilon U_{1}(\rho,\psi)~,\\
    V(\rho,\psi)&=&\epsilon V_{1}(\rho,\psi)~.
\end{eqnarray}
\end{subequations}
The first step is to use a
gauge transformation of the form
$\rho\rightarrow\rho(1-\epsilon F_{1})$,
$\psi\rightarrow\psi+\epsilon F_{1}\tan\psi$, which leaves the last term of (\ref{spherical}) unchanged to leading order, to
choose a gauge where
\begin{equation}
    V_{1}(\rho,\psi)=\frac{\rho(2\rho^{2}-\mu^{2})}
    {4(\rho^{2}-\mu^{2})}A_{1,\psi}(\rho,\psi)
    +\frac{\rho^{2}\cot\psi}{2}U_{1,\rho}(\rho,\psi)~.
    \label{gauge}
\end{equation}
This gauge choice simplifies the equations for the rest of the functions to be determined.

Next, one can solve the equation
$G_{\rho\psi}=0$ for $B_{1}$. The rest of the equations can be
solved in terms of a single function,  $H_1(\rho,\psi)$, obeying a second order partial
differential equation. The solution is
\begin{subequations}
\label{N1solution}
\begin{eqnarray}
    B_{1}&=&-\frac{2\mu^{2}}{\rho^{2}}F_{1}(\rho,\psi)~,
    \label{BN1}\\
    A_{1}&=&\frac{4H_{1,\psi}(\rho,\psi)}{\sin^{2}\psi\,\cos\psi}
    +2F_{1}(\rho,\psi)+2\rho F_{1,\rho}(\rho,\psi)~,
    \label{AN1}\\
    U_{1}&=&-2
    \frac{6\rho H_{1}(\rho,\psi)
    -(2\rho^{2}-\mu^{2})H_{1,\rho}(\rho,\psi)}{\rho\sin\psi\,\cos^{2}\psi}
    -2\tan^{2}\psi F_{1}(\rho,\psi)-2\tan\psi F_{1,\psi}(\rho,\psi)~,
    \label{UN1}\\
    V_{1}&=& \frac{\rho(2\rho^{2}-\mu^{2})}{(\rho^{2}-\mu^{2})\sin^{2}\psi\,\cos\psi}
    \left[3H_{1}(\rho,\psi)-\tan\psi
    H_{1,\psi}(\rho,\psi)\right]\nonumber\\& &
    -\frac{2\rho^{2}H_{1,\rho}(\rho,\psi)}{\sin^{2}\psi\,\cos\psi}
    +\frac{\rho^{3}}{\rho^{2}-\mu^{2}}F_{1,\psi}(\rho,\psi)
    -\rho^{2}\tan\psi F_{1,\rho}(\rho,\psi)
    ~.
    \label{VN1}
\end{eqnarray}
\end{subequations}
The wave function $H_{1}(\rho,\psi)$ satisfies the differential equation
\begin{equation}
    (\rho^{2}-\mu^{2})\left(H_{1,\rho\rho}-\frac{1}{\rho}H_{1,\rho}\right)+H_{1,\psi\psi}
    -4\cot2\psi H_{1,\psi}-3H_{1}=0~.
    \label{H1equation}
\end{equation}
The solution of Eq.(\ref{H1equation}) can be written as
\begin{eqnarray}
    H_{1}(\rho=R\mu,\psi)&=&2R\sqrt{R^{2}-1}\int\,d\nu
    \left[a(\nu)
        P^{1}_{\nu}(2R^{2}-1)+b(\nu)
        Q^{1}_{\nu}(2R^{2}-1)\right]h_{\nu}(\psi)~,
    \label{H1nu}\\
    h_{\nu}(\psi)&=&\frac{4}{\sqrt{2\pi}}\sin^{3/2}2\psi\left[
    \pi\cos\alpha_{\nu}P^{3/2}_{\nu}(\cos2\psi)
    -2\sin\alpha_{\nu}Q^{3/2}_{\nu}(\cos2\psi)\right]\nonumber\\
    & &=(2\nu-1)\cos[(2\nu+3)\psi-\alpha_{\nu}]
    -(2\nu+3)\cos[(2\nu-1)\psi-\alpha_{\nu}]~,
    \label{hpsi}
\end{eqnarray}
where $P^{\mu}_{\nu}$ and $Q^{\mu}_{\nu}$ are associated Legendre
functions of the first and second kind. The functions $a(\nu)$ and $b(\nu)$ will be fixed below using the constraints on function $H_{1}$. These constraints come from
symmetries, from regularity requirements, and from boundary
conditions.

\subsubsection{ Conditions at $\psi=\pi/2$} $Z_{2}$ symmetry about the
direction
   $\psi=\pi/2$ ($w=0$) implies the conditions
\begin{subequations}
\label{z2brane}
\begin{eqnarray}
    H_{1,\psi}(\rho,\pi/2)&=&0~,\label{H1,psi}\\
    F_{1,\psi}(\rho,\pi/2)&=&0~,\label{F1,psi}\\
    F_{1}(\rho,\pi/2)&=&-6H_{1}(\rho,\pi/2)+
    \frac{2\rho^{2}-\mu^{2}}{\rho}H_{1,\rho}(\rho,\pi/2)~.
    \label{F1brane}
\end{eqnarray}
\end{subequations}
It can be verified that condition (\ref{H1,psi}) is automatically satisfied by (\ref{H1nu}).

\subsubsection{Conditions at $\psi=0$} For a black hole configuration, unlike for a black string configuration,  the mass is localized at the origin.  Then  the components of the metric, (\ref{N1}),
are finite at $\psi=0$ and arbitrary $\rho$. This will only be true if
\begin{equation}
    H_{1}(\rho,0)=H_{1,\psi}(\rho,0)=H_{1,\psi\psi}(\rho,0)=0
    \label{nostring}
\end{equation}
It can be verified that  (\ref{nostring}) implies 
$\cos\alpha_{\nu}=0$. Then  (\ref{hpsi}) simplifies to
\begin{equation}
    h_{\nu}(\psi)=(2\nu-1)\sin(2\nu+3)\psi
    -(2\nu+3)\sin(2\nu-1)\psi~,
    \label{hpsisin}
\end{equation}
where we have omit a possible overall sign.

\subsubsection{Evenness in $\mu$} As we mentioned earlier,
the expansion in $\epsilon=(\mu/L)^{2}$ and the fact that
the zeroth order of the near solution (\ref{N0}) depends only
on $\mu^{2}$ imply that the metric is even in $\mu$.
This means that the
function $H_{1}$ should be even in $R=\rho/\mu$. For large $R$ the
Legendre functions in (\ref{H1nu}) behave as
\begin{eqnarray}
    R\sqrt{R^{2}-1}P_{\nu}^{1}(2R^{2}-1)&\sim&R^{2\nu+2}\times f_{P}(R^{-2})~,\\
    R\sqrt{R^{2}-1}Q_{\nu}^{1}(2R^{2}-1)&\sim&R^{-2\nu}\times f_{Q}(R^{-2})~,
\end{eqnarray}
where $f_{P/Q}$ are analytic functions. Since we require that $H_{1}$ is even in $R$,
$\nu$ should be integer. The integral in (\ref{H1nu}) is reduced to a sum over
integer values of $\nu$.
\begin{eqnarray}
    H_{1}(\rho=R\mu,\psi)&=&\left[a_{0}(R^{2}-1)+b_{0}\right]\sin^{3}\psi
    +2R\sqrt{R^{2}-1}\sum_{n=1}^{\infty}
    \left[a_{n}
        P^{1}_{n}(2R^{2}-1)+b_{n}
        Q^{1}_{n}(2R^{2}-1)\right]\nonumber\\
    & &\hspace{2in}\times\left[(2n-1)\sin(2n+3)\psi
    -(2n+3)\sin(2n-1)\psi\right]~.
    \label{H1sum}
\end{eqnarray}
The case $n=0$ requires special attention, since the $  P^{1}_{0}\equiv0$.

\subsubsection{Requirement of a Killing horizon} The metric (\ref{spherical}) is static.
Therefore, the surface $B=0$ is a Killing horizon. The normal
vector $B_{,\alpha}$ should be null on the horizon. This implies that
on the horizon $B_{,\psi}=0$, thus the horizon is located at
constant $\rho=\rho_{H}$. Using  metric (\ref{N1}) we
expand $\rho_{H}$ in $\epsilon$ as
$\rho_{H}=\mu(1+\epsilon\zeta_{1})$. The  conditions
$B(\rho_{H},\psi)=0$ and $B_{,\psi}(\rho_{H},\psi)=0$ restrict the
gauge function $F_{1}$
\begin{equation}
    F_{1}(\rho=\mu,\psi)=-\zeta_{1},\;\;\;F_{1,\psi}(\rho=\mu,\psi)=0~.
    \label{F1horizon}
\end{equation}
The surface gravity for metric (\ref{spherical}) is defined as
\begin{equation}
    \kappa=\left.\frac{B_{,\rho}(\rho,\psi)}{2\sqrt{A(\rho,\psi)}}\right|_{\rho=\rho_{H}}~.
    \label{kappa}
\end{equation}

The surface gravity should be constant on the horizon, otherwise the horizon is
singular \cite{wald}. When we expand the metric  in $\epsilon$,
the surface gravity must be constant in every order of the expansion.
We use  metric (\ref{N1}) and  conditions (\ref{F1horizon})
to evaluate the surface gravity in order $\epsilon$
\begin{equation}
    \kappa=1-2\epsilon\frac{H_{1,\psi}(\mu,\psi)}{\sin^{2}\psi\,\cos\psi}
    =1+\epsilon\chi_{1}~.
    \label{kappaH1}
\end{equation}
The requirement of a constant surface gravity constrains 
the function $H_{1}$. At $\rho=\mu$ the Legendre functions in
Eq.(\ref{H1sum}) take the values
$\left.2R\sqrt{R^{2}-1}P_{n}^{1}(2R^{2}-1)\right|_{R=1}=0$,
$\left.2R\sqrt{R^{2}-1}Q_{n}^{1}(2R^{2}-1)\right|_{R=1}=-1$. We
use the representation (\ref{H1sum}) to evaluate the surface
gravity (\ref{kappaH1})
\begin{equation}
    -6b_{0}-4\sum_{n=1}^{\infty}b_{n}(2n-1)(2n+3)\frac{\sin(2n+1)\psi}{\sin\psi}
    =\chi_{1}=\text{constant}~.
    \label{chi1}
\end{equation}
The set $\left\{\sin(2n+1)\psi\right\}$ is complete on the
interval $[0,\pi/2]$, therefore, the solution to Eq.(\ref{chi1})
is
\begin{equation}
    -6b_{0}=\chi_{1},\;\;\;b_{n>0}=0~.
    \label{bn}
\end{equation}
In other words, the sum in Eq.(\ref{H1sum}) contains Legendre functions of the first kind only.
The functions $2R\sqrt{R^{2}-1}P_{n}^{1}(2R^{2}-1)$
are polynomials of order $n+1$ in $(2R^{2}-1)$.

\subsection{Matching the near and the asymptotic solutions}
The remaining free parameters of the near solution, $\{a_{n},b_{0},\zeta_{1}\}$, must be determined from matching  the two gauge functions, $W(r,w)$ in the asymptotic solution  (\ref{A1}) and $F_{1}(\rho,\psi)$
in the near solution (\ref{N1}).
We use (\ref{transformation}) to transform the asymptotic
solution to $(\rho,\psi)$ coordinates and then we expand it in $L^{-1}$.
The gauge function $W(r,w)$ is also expanded as
\begin{equation}
    W(r,w)=\cos\frac{\pi w}{L}\sum_{n=0}^{\infty}
    \left(\frac{\pi}{L}\right)^{2n}W_{2n}(\rho,\psi)~.
    \label{Wn}
\end{equation}
We keep the explicit factor $\cos\frac{\pi w}{L}$ in (\ref{Wn}) to insure that
 $W(r,L/2)=0$.

\subsubsection{ Matching in zeroth order} We start with zeroth order in
$L^{-1}$. We transform the asymptotic metric using
(\ref{transformation}) and expand it to zeroth order in $L^{-1}$.
\begin{subequations}
\label{asymprhopsi0}
\begin{eqnarray}
    B^{A}&=&1-\frac{\mu^{2}}{\rho^{2}}~,
    \label{Basymp0}\\
    A^{A}&=&1-\frac{\mu^{2}\cos\psi(\cos\psi-2\rho^{2}W_{0,\rho}(\rho,\psi))}
    {\rho^{2}}~,
    \label{Aasymp0}\\
    U^{A}&=&1+\frac{\mu^{2}\sin\psi(\sin\psi-2\rho W_{0,\psi}(\rho,\psi))}
    {\rho^{2}}~,
    \label{Uasymp0}\\
    V^{A}&=& \mu^{2}\left[\cos\psi W_{0,\psi}(\rho,\psi)
    -\rho\sin\psi W_{0,\rho}(\rho,\psi)\right]~,
    \label{Vasymp0}
\end{eqnarray}
\end{subequations}
where the superscript $A$ stands for 'Asymptotic' and the functions are
defined in ansatz (\ref{spherical}).
Matching (\ref{asymprhopsi0}) to zeroth order of the near solution (\ref{N0})
determines the function $W_{0}$
\begin{equation}
    W_{0}(\rho,\psi)=-\frac{\cos\psi}{2\rho}~.\label{W0}
\end{equation}

\subsubsection{Matching in $O(L^{-2})$ } The asymptotic solution is
\begin{subequations}
\label{asymprhopsi1}
\begin{eqnarray}
    B^{A}&=&1-\frac{\mu^{2}}{\rho^{2}}-\frac{\mu^{2}\pi^{2}
    }{3L^{2}}(1-\cos^{4}\psi)~,
    \label{Basymp1}\\
    A^{A}&=&1+\frac{\pi^{2}}{L^{2}}\left[\rho^{2}\cos^{4}\psi
    +\frac{\mu^{2}(41\cos2\psi+8\cos4\psi-\cos6\psi)}{96}
    +2\mu^{2}\cos\psi W_{2,\rho}(\rho,\psi)\right]~,
    \label{Aasymp1}\\
    U^{A}&=&1+\frac{\pi^{2}}{L^{2}}\left[\rho^{2}\cos^{2}\psi\sin^{2}\psi
    -\frac{\mu^{2}\sin\psi\cos^{2}\psi(5\sin3\psi-37\sin\psi)}{12}
    -2\mu^{2}\frac{\sin\psi}{\rho} W_{2,\psi}(\rho,\psi)\right]~,
    \label{Uasymp1}\\
    V^{A}&=& -\frac{\pi^{2}}{L^{2}}\left[\rho^{3}\sin\psi\cos^{3}\psi
    +\mu^{2}\rho\sin\psi\cos^{3}\psi(\sin^{2}\psi+2)
    +\mu^{2}\rho\sin\psi W_{2,\rho}(\rho,\psi)
    -\mu^{2}\cos\psi W_{2,\psi}(\rho,\psi)\right] ~.
    \label{Vasymp1}
\end{eqnarray}
\end{subequations}
The near solution (\ref{N1}) should be expanded to second order in
$\mu$, which appears in $\epsilon=(\mu/L)^{2}$ and in the rescaled
coordinate $R=\rho/\mu$. The large $R$ behavior of the terms of $H_{1}$, (\ref{H1sum}), is
\begin{equation}
    2R\sqrt{R^{2}-1}P_{n}^{1}(2R^{2}-1)\sim
    \frac{2(2n)!}{n!(n-1)!}R^{2n+2}~.
\end{equation}
These expressions contribute by terms of
order $\epsilon H_{1}\sim L^{-2}\mu^{-2n}$ in metric (\ref
{spherical}). We deduce that $n>0$ terms contribute by negative orders in
$\mu$ and must be eliminated. Therefore we impose $a_{n}=0$ if $n>0$. 
For similar
reasons,  gauge function $F_{1}$ should also be a polynomial
of order $2$ in $\rho$. Thus,  functions $H_{1}$ and $F_{1}$ are
\begin{eqnarray}
    H_{1}&=&[a_{0}(\mu^{-2}\rho^{2}-1)+b_{0}]\sin^{3}\psi~,\label{H10}\\
    F_{1}&=&(\mu^{-2}\rho^{2}-1)f_{2}(\psi)-\zeta_{1}~,\label{F1asymp}
\end{eqnarray}
where we have already imposed $F_{1}(\rho=\mu,\psi)=-\zeta_{1}$, (\ref{F1horizon}). The exact form of the periodic function $f_2(\psi)$ will be fixed below.
Then the near metric, expanded to second order in $\mu$, is
\begin{subequations}
\label{nearmu2}
\begin{eqnarray}
    B^{N}&=&1-\frac{\mu^{2}}{\rho^{2}}+\frac{2\mu^{2}}{L^{2}}f_{2}(\psi)~,
    \label{Bnear2}\\
    A^{N}&=&1+\frac{1}{L^{2}}\left[12a_{0}\rho^{2}+6\rho^{2}f_{2}(\psi)
    -2\mu^{2}(6a_{0}-6b_{0}+f_{2}(\psi)+\zeta_{2})\right]~,
    \label{Anear2}\\
    U^{N}&=&1-\frac{2\tan^{2}\psi}{L^{2}}\left[\rho^{2}[2a_{0}+f_{2}(\psi)
    +\cot\psi f_{2}'(\psi)]-\mu^{2}[4a_{0}-6b_{0}+f_{2}(\psi)+\zeta_{2}
    +\cot\psi f_{2}'(\psi)]\right]~,
    \label{Unear2}\\
    V^{N}&=& \frac{\rho^{3}\tan\psi}{L^{2}}
    \left[-4a_{0}-2f_{2}(\psi)+\cot\psi f_{2}'(\psi)\right]    ~.
    \label{Vnear2}
\end{eqnarray}
\end{subequations}
Comparing (\ref{asymprhopsi1}) and (\ref{nearmu2}) we find that
\begin{eqnarray}
    f_{2}(\psi)&=&-\frac{\pi^{2}}{12}(3+\cos2\psi)\sin^{2}\psi~,\label{f2}\\
    a_{0}&=&\frac{\pi^{2}}{12}~,\label{a0}\\
    W_{2}(\rho,\psi)&=&\frac{\rho}{192\cos\psi}\left[
    -76+1152\pi^{-2}b_{0}-57\cos2\psi-12\cos4\psi+\cos6\psi-192\pi^{-2}\zeta_{1}\right]
    ~.\label{W2}
\end{eqnarray}
At this point the near solution still contains two free parameters, $b_{0}$
and $\zeta_{1}$.
The $Z_{2}$-symmetry condition, (\ref{F1brane}), imposes one constraint on these
parameters
\begin{equation}
    b_{0}=\frac{\pi^{2}+6\zeta_{1}}{36}~.\label{b0}
\end{equation}
A constraint, determining  $\zeta_{1}$, is derived from the first law of black hole thermodynamics.
It will be discussed in the next section.

\subsection{Black hole thermodynamics}
The zeroth law of black hole thermodynamics states that
the temperature of a black hole is
\begin{equation}
    T=\frac{\kappa}{2\pi}~,\label{T}
\end{equation}
where $\kappa$ is the surface gravity, which is constant on the
horizon. To calculate the temperature for the near solution we use
Eqs.(\ref{kappa}, \ref{kappaH1}, \ref{bn}) to find that
\begin{equation}
    T=\frac{1-6b_{0}\epsilon}{2\pi\mu}~.\label{T1}
\end{equation}

The first law of black hole thermodynamics is $dM=TdS$, where the
entropy is proportional to the area of the horizon
\begin{equation}
    S\equiv\frac{A_{H}}{4G_{5}}=\frac{1}{4G_{5}}
    \int_{0}^{\pi}d\psi\int_{0}^{\pi}d\theta
    \int_{0}^{2\pi}d\phi\left.\sqrt{g_{\psi\psi}g_{\theta\theta}g_{\phi\phi}}
    \right|_{\rho=\rho_{H}}~.
    \label{horizonarea}
\end{equation}
We calculate the entropy for the near solution and find that
\begin{equation}
    S=\frac{\pi^{2}\mu^{3}(1+3\zeta_{1}\epsilon)}{2G_{5}}~.
    \label{S}
\end{equation}
According to the first law the temperature can be found as
$T^{-1}=\partial S/\partial M$. The mass appears in the
entropy only through $\mu=\sqrt{8G_{5}M/(3\pi)}$ and
$\epsilon=(\mu/L)^{2}$, so if we combine the first law and
the zeroth law we get
\begin{equation}
    T^{-1}=2\pi\mu[1+6b_{0}\epsilon]=2\pi\mu[1+5\zeta_{1}\epsilon]~.
    \label{T2}
\end{equation}
This fixes the last parameter. If we combine Eqs.(\ref{b0},\ref{T2}) we find that
\begin{equation}
    \zeta_{1}=\frac{\pi^{2}}{24};\;\;\;b_{0}=\frac{5\pi^{2}}{144}~.
\end{equation}
To summarize, the near metric, to first order in $\epsilon$, is
\begin{subequations}
\label{near1final}
\begin{eqnarray}
    B^{N}&=&1-\frac{\mu^{2}}{\rho^{2}}-\frac{\pi^{2}\epsilon}{12\rho^{2}}\left[
    4(1-\cos^{4}\psi)(\rho^{2}-\mu^{2})+\mu^{2}\right]~,
    \label{B1final}\\
    A^{N}&=&1+\frac{\pi^{2}\epsilon}{3}\left[(3\frac{\rho^{2}}{\mu^{2}}-1)\cos^{4}\psi-1
    \right]~,
    \label{A1final}\\
    U^{N}&=&1+\pi^{2}\epsilon(\frac{\rho^{2}}{\mu^{2}}-1)\sin^{2}\psi\cos^{2}\psi~,
    \label{U1final}\\
    V^{N}&=& -\pi^{2}\epsilon\frac{\rho^{3}}{\mu^{2}}\sin\psi\cos^{3}\psi   ~.
    \label{V1final}
\end{eqnarray}
\end{subequations}
The location of the horizon is at
$\rho_{H}=\mu(1+\pi^{2}\epsilon/24)$. The entropy and the
temperature are
\begin{eqnarray}
    S&=&\frac{\pi^{2}\mu^{3}}{2G_{5}}\left(1+\frac{\pi^{2}\mu^{2}}{8L^{2}}\right)~,\label{S1final}\\
    T&=&\frac{1}{2\pi\mu}\left(1-\frac{5\pi^{2}\mu^{2}}{24L^{2}}\right)~.\label{T1final}
\end{eqnarray}
These expressions are in agreement with previous results~\cite{harmark2,gorbonos}.
The only freedom left in the first order metric is the $n>1$ terms of the gauge function
 in the asymptotic solution (\ref{Wn}). The $n=0$ and $n=2$ terms are fixed in the region $\rho\ll L$
by our matching procedure, as given in Eqs.(\ref{W0},\ref{W2}).
In addition, in the asymptotic region, $r\gg L$, the metric should be Minkowski, therefore,
we require that $W(r\gg L,w)={\cal O}(r^{1})$.
A form of the gauge function which is consistent with these conditions appears
in appendix \ref{app:asymp2}.

\section{Higher order corrections}
\label{sec:high}
  Prior to completing calculations in second order we describe our general procedure
for calculating higher order contributions. First we consider the asymptotic
solution, which is fully determined (up to gauge freedom) by the lower order contributions.
\subsection{The $n$th order asymptotic solution}
In the asymptotic region we expand the metric in $\epsilon$ as
\begin{equation}
    g_{\mu\nu}=\eta_{\mu\nu}+\sum_{n=1}^{\infty}\epsilon^{n}h_{\mu\nu}^{(n)}(r,w)~.
    \label{An}
\end{equation}
Assume that we know the solution up to order $(n-1)$ and we intend to obtain the
$n$th order solution. Einstein's equation is linear in the $h^n_{\mu\nu}$ but 
due to lower order contributions it is inhomogeneous. The solution is similar to (\ref{h1}) but, in addition to the solution of the homogeneous equation
includes extra terms corresponding to a particular solution
(the particular solution is denoted by $\bar{h}_{\mu\nu}$), as follows
\begin{subequations}
\label{hn}
\begin{eqnarray}
    h_{rr}^{(n)}&=&\bar{h}_{rr}^{(n)}(r,w)+\frac{\alpha_n}{r}-\frac{r}{2}h_{tt,r}^{(n)}~,\\
    h_{ww}^{(n)}&=&\bar{h}_{ww}^{(n)}(r,w)-\frac{\alpha_n}{r}
    -\frac{r}{2}h_{tt,r}^{(n)}+2W_{,w}^{(n)}(r,w)~,\\
    h_{rw}^{(n)}&=&\bar{h}_{rw}^{(n)}(r,w)+W_{,r}^{(n)}(r,w)~,
\end{eqnarray}
\end{subequations}
where $\alpha$ is a constant, and $W^{(n)}(r,w)$ is a gauge function which is periodic and
antisymmetric in $w$.
The function $h_{tt}^{(n)}$ is also periodic and symmetric in $w$. It satisfies the equation
\begin{equation}
    h_{tt,rr}^{(n)}+\frac{2}{r}h_{tt,r}^{(n)}+h_{tt,ww}^{(n)}=S^{(n)}~,
    \label{httnEq}
\end{equation}
where $S^{(n)}$ is the deriving term, which depends on lower orders, $h^{(k)}_{\mu\nu}$, $k<n$.
The homogeneous solution to Eq.(\ref{httnEq}) can be written as a Fourier series. Consequently, we have
\begin{equation}
    h_{tt}^{(n)}(r,w)=\bar{h}_{tt}^{(n)}(r,w)+\sum_{k=0}^{\infty}c_{k}\frac{\exp(-2\pi k r/L)}{r}
    \cos(2\pi k w/L)~.
    \label{httnmodes}
\end{equation}
The particular solution $\bar{h}_{\mu\nu}$ is completely determined by the lower orders of
the asymptotic solution (without any use of the near solution).
The free parameters are $\alpha$ and the set $\{c_{k}\}$.
These can be determined by matching to the lower orders of the near solution
as follows.
Take $g_{tt}$ from the near solution up to order $\epsilon^{n-1}$ and expand it in
$\mu$. Take the term of order $\mu^{2n}$ and find its Fourier series just like in Eq.(\ref{Fourier1}).
Compare that Fourier series with the Forier series of $h_{tt}^{(n)}$ to lowest order in $L^{-1}$,
and determine the constants $\{c_{k}\}$.

Just like in first order, we can show that  $\alpha_n=0$. Solution (\ref{hn}) should be even in $L$. The functions $\bar{h}^{(n)}$
and $h_{tt}^{(n)}$ are even functions of $L$ since these functions are determined by
lower orders but for dimensional reasons $\alpha_n\sim L$ is odd. Therefore,   $\alpha_n$ must vanish.

Next, we have to make sure that the definition of the ADM mass does not change.
So, we require that
\begin{equation}
    \lim_{r\rightarrow\infty}r^{2}\int_{-L/2}^{L/2}
    \left[\frac{2}{r}h_{rr}^{(n)}-h_{ww,r}^{(n)}\right]dw=0~.
    \label{fixM}
\end{equation}
In general, the tension in the compact dimension, (\ref{tension}), does not vanish.
As a result, the effective four-dimensional Newton's constant, $G_{4}$, acquires a correction
of order $\epsilon^{n}$. This means that $G_{4}$ depends on the mass and the {\em equivalence
principle is violated.}

At this point the asymptotic solution is determined up to order $\epsilon^{n}$,
except for the gauge function $W^{(n)}(r,w)$, which is (partially) determined by matching
to the near solution to orders up to $\epsilon^{n}$.

\subsection{The $n$th order near solution}
\label{subsec:nearn} Near the horizon  it is convenient
to use the metric in form (\ref{spherical}) and expand it
in $\epsilon$ as follows
\begin{subequations}
\label{emetric}
\begin{eqnarray}
    B(\rho,\psi)&=&1-\frac{\mu^{2}}{\rho^{2}}-\sum_{n=1}\epsilon^{n} B_{n}(\rho,\psi)~,\\
    A(\rho,\psi)&=&1+\sum_{n=1}\epsilon^{n} A_{n}(\rho,\psi)~,\\
    U(\rho,\psi)&=&1+\sum_{n=1}\epsilon^{n} U_{n}(\rho,\psi)~,\\
    V(\rho,\psi)&=&\sum_{n=1}\epsilon^{n} V_{n}(\rho,\psi)~.
\end{eqnarray}
\end{subequations}
 We need the near
solution up to order $(n-1)$ and the asymptotic solution up to
order $n$ to calculate the $n$th order near solution. Again, the $n$th order Einstein's equation is a linear
inhomogeneous equation in
$g^{(n)}\equiv\{B_{n},A_{n},U_{n},V_{n}\}$. The inhomogeneity 
depends on $g^{(k<n)}$. We solve the equations, in a way,
similar to that for the first order correction. The first step is to
apply a gauge transformation of the form
$\rho\rightarrow\rho(1-\epsilon^{n}F_{n})$,
$\psi\rightarrow\psi+\epsilon^{n}F_{n}\tan\psi$ to arrive at a gauge, in which
\begin{equation}
    V_{n}(\rho,\psi)=\frac{\rho(2\rho^{2}-\mu^{2})}{4(\rho^{2}-\mu^{2})}A_{n,\psi}(\rho,\psi)
    +\frac{\rho^{2}\cot\psi}{2}U_{n,\rho}(\rho,\psi)~.
    \label{gaugen}
\end{equation}
Next, one can solve the inhomogeneous equation
$G_{\rho\psi}=0$ for $B_{n}$. The rest of the equations can be
solved in terms of a single function, $H_{n}(\rho,\psi)$,
obeying a second order partial
differential equation, which is similar to (\ref{H1equation}),
\begin{equation}
    (\rho^{2}-\mu^{2})\left(H_{n,\rho\rho}-\frac{1}{\rho}H_{n,\rho}\right)+H_{n,\psi\psi}
    -4\cot2\psi H_{n,\psi}-3H_{n}=S_{n}~,
    \label{Hnequation}
\end{equation}
where $S_{n}$ depends on the lower order corrections. The solution
of Eq.(\ref{Hnequation}) can be found by the method of separation of
variables, just like it has been done when we have calculated the first order correction (\ref{H1nu}). The
eigenfunctions of Eq.(\ref{Hnequation}) are the functions
$h_{\nu}(\psi)$, which appear in Eq.(\ref{hpsi}). The driving
term, $S_{n}$, can also be expanded in the set $h_{\nu}(\psi)$. In
addition, the boundary conditions at $\psi=\pi/2$ and at $\psi=0$
in $n$th order are the same as in first order,
Eqs.(\ref{H1,psi}, \ref{nostring}). And just like in first order
 the function $H_{n}$ should be even in $\rho$. So, the
general solution of Eq.(\ref{Hnequation}) is
\begin{eqnarray}
    H_{n}(R=\rho/\mu,\psi)&=&\left[a_{0}(R^{2}-1)+b_{0}+p_{0}(R)\right]\sin^{3}\psi\nonumber\\
    & &+\sum_{n=1}^{\infty}
    \left[2R\sqrt{R^{2}-1}\left(a_{n}
        P^{1}_{n}(2R^{2}-1)+b_{n}
        Q^{1}_{n}(2R^{2}-1)\right)+p_{n}(R)\right]\nonumber\\
    & &\hspace{2in}\times\left[(2n-1)\sin(2n+3)\psi
    -(2n+3)\sin(2n-1)\psi\right]~,
    \label{Hnsum}
\end{eqnarray}
where $\left\{p_{n}(R)\right\}$ is a particular solution of
the inhomogeneous  equation (\ref{Hnequation}).
The particular solutions are completely determined by the lower orders of the
near solution (without using the asymptotic solution).
The homogeneous part  should be completely fixed by matching to the
asymptotic solution. The matching is done as follows: Take the asymptotic
solution up to order $\epsilon^{n}$ and transform it to the $(\rho,\psi)$
coordinates, using  (\ref{transformation}).
Expand the solution in $L^{-1}$, take the term of order $L^{-2n}$ and
compare it to the near solution of order $\epsilon^{n}$ to fix the
parameters $\{a_{n},b_{n}\}$, and the gauge functions $F_{n}(\rho,\psi)$
and $W^{(n)}(r,w)$. At this point the function $W^{(n)}(r,w)$ is determined
only up to order $L^{-2n}$ and one should fix it completely before proceeding to
the next order. In the next section we apply this method to calculate the
second order contributions.

\section{Second order in $\epsilon$}
\label{sec:second}
\subsection{The asymptotic solution - second order}
Up to first order, the asymptotic solution  is given by equations
(\ref{A0}), (\ref{A1}), (\ref{h1}), and (\ref{htt1final}).
The second order solution is described in the  Appendix \ref{app:asymp2}.
It contains an inhomogeneous part, $\bar{h}^{(2)}(r,w)$, which is
completely determined by the first order solution,
and a homogeneous part,
which includes the wave function $h_{tt}^{(H)}(r,w)$, and the gauge function
$W^{(2)}(r,w)$
\begin{subequations}
\label{h2}
\begin{eqnarray}
    h_{tt}^{(2)}&=&h_{tt}^{(H)}+\bar{h}^{(2)}_{tt}~,\\
    h_{rr}^{(2)}&=&-\frac{r}{2}h_{tt,r}^{(H)}+\bar{h}^{(2)}_{rr}~,\\
    h_{ww}^{(2)}&=&-\frac{r}{2}h_{tt,r}^{(2)}+2W_{,w}^{(2)}+\bar{h}_{ww}^{(2)}~,\\
    h_{rw}^{(2)}&=&W_{,r}^{(2)}+\bar{h}_{rw}^{(2)}~.
\end{eqnarray}
\end{subequations}
The wave function $h_{tt}^{(H)}$ satisfies the homogeneous part of Eq.(\ref{httnEq}),
and can be written in the form of Eq.(\ref{httnmodes})
\begin{equation}
    h_{tt}^{(H)}(r,w)=\sum_{k=0}^{\infty}c_{k}\frac{\exp(-2\pi k r/L)}{r}
    \cos(2\pi k w/L)~.
    \label{htt2modes}
\end{equation}
To fix the constants, $\{c_{k}\}$, we follow the prescription that appears
after Eq.(\ref{httnmodes}). We take $g_{tt}$ from the first order near solution,
(\ref{B1final}), and expand it in $\mu$, up to order $\mu^{4}$, to get
\begin{equation}
    g_{tt}^{N}=-1+\mu^{2}\left[\frac{1}{\rho^{2}}
    +\frac{\pi^{2}(1-\cos^{4}\psi)}{3L^{2}}\right]
    +\mu^{4}\frac{\pi^{2}(4\cos^{4}\psi-3)}{12L^{2}\rho^{2}}~.
    \label{gttN1}
\end{equation}
Then we take $g_{tt}$ from the second order asymptotic solution and transform it  to the $(\rho,\psi)$ coordinates, using
(\ref{transformation}). After expanding it in $L^{-1}$,
up to order $L^{-2}$, we obtain
\begin{equation}
    g_{tt}^{A}=-1+\mu^{2}\left[\frac{1}{\rho^{2}}
    +\frac{\pi^{2}(1-\cos^{4}\psi)}{3L^{2}}\right]
    +\mu^{4}\left[\frac{2\cos2\psi+1}{4\rho^{4}}
    -\frac{\pi^{2}(3\cos6\psi+124\cos^{4}\psi+21\cos2\psi+4)}
    {96L^{2}\rho^{2}}\right]+\epsilon^{2}h_{tt}^{(H)}~.
    \label{gttA2}
\end{equation}
We find $h_{tt}^{(H)}$ from $g_{tt}^{A}-g_{tt}^{N}$  and calculate its Fourier 
coefficients
\begin{eqnarray}
    h_{tt}^{(H)}&=&-\frac{\pi^{3}L}{12r}\left[1-2\sum_{k=1}^{\infty}(6k^{2}-1)
    e^{-2\pi k r/L}\cos\frac{2\pi k w}{L}\right]\nonumber\\
    & &=\frac{\pi^{3}L\sinh\frac{2\pi r}{L}\left(-\cosh\frac{4\pi r}{L}
    +16\cosh\frac{2\pi r}{L}\cos\frac{2\pi w}{L}+5\cos\frac{4\pi w}{L}-20\right)}
    {24r\left(\cosh\frac{2\pi r}{L}-\cos\frac{2\pi w}{L}\right)^{3}}~.
    \label{httH}
\end{eqnarray}
The asymptotic ($r\gg L$) form of the metric is
\begin{subequations}
\label{h2asymp}
\begin{eqnarray}
    h_{tt}^{(2)}&\sim&-\frac{\pi^{3} L}{12r}~,\\
    h_{rr}^{(2)}&\sim&\frac{\pi^{3}L}{12r}~,\\
    h_{ww}^{(2)}&\sim&\frac{\pi^{3}L}{12r}+2W_{,w}^{(2)}~,\\
    h_{rw}^{(2)}&\sim&W_{,r}^{(2)}~.
\end{eqnarray}
\end{subequations}
The contribution to the ADM mass  vanishes indeed,
\begin{equation}
    \Delta M\propto\lim_{r\rightarrow\infty}r^{2}\int_{-L/2}^{L/2}
    \left[\frac{2}{r}h_{rr}^{(2)}-h_{ww,r}^{(2)}\right]dw
    =\frac{\pi^{3}L^{2}}{4}-2r^{2}\left.W_{,r}^{(2)}\right|_{w=-L/2}^{w=L/2}=0
    ~,
    \label{DeltaM2}
\end{equation}
where the last equality follows from  boundary condition (\ref{W2L2}).
The tension (\ref{tension}) is now
\begin{equation}
    b=\int d^{4}x T^{ww}=-\frac{\epsilon^{2}}{4G_{5}}\lim_{r\rightarrow\infty}r^{2}
        \int_{-L/2}^{L/2}\left[\frac{2}{r}h_{rr}+h_{tt,r}\right]dw
        =-M\frac{\pi^{2}\epsilon}{6}~.
    \label{tension2}
\end{equation}

\subsection{Second order correction to the near solution}
Following section \ref{subsec:nearn}, we find that the second order solution
is
\begin{subequations}
\label{emetric2}
\begin{eqnarray}
    B_{2}&=&-\frac{2\mu^{2}}{\rho^{2}}F_{2}(\rho,\psi)
    +\frac{\pi^{4}\cos^{4}\psi}{216\mu^{2}\rho^{2}}
    \left[3(\rho^{4}-\mu^{2}\rho^{2}+\mu^{4})\cos4\psi\right.\nonumber\\&
    & \left.
    +4(4\rho^{4}-6\mu^{2}\rho^{2}+3\mu^{4})\cos2\psi-
    5(7\rho^{4}-21\mu^{2}\rho^{2}-3\mu^{4})\right]~,
    \label{B2}\\
    A_{2}&=&\frac{4H_{2,\psi}(\rho,\psi)}{\sin^{2}\psi\,\cos\psi}
    +2F_{2}(\rho,\psi)+2\rho F_{2,\rho}(\rho,\psi)-\frac{\pi^{4}\cos^{2}\psi}{864\mu^{4}}
    \left[\mu^{2}(105\rho^{2}+11\mu^{2})\cos6\psi\right.\nonumber\\
    & &\left.+(-64\rho^{4}+138\mu^{2}\rho^{2}+58\mu^{4})\cos4\psi
    +(1472\rho^{4}+1383\mu^{2}\rho^{2}+237\mu^{4})\cos2\psi
    -768\rho^{4}+774\mu^{2}\rho^{2}-2\mu^{4}\right]~,
    \label{A2}\\
    U_{2}&=&-2
    \frac{6\rho H_{2}(\rho,\psi)
    -(2\rho^{2}-\mu^{2})H_{2,\rho}(\rho,\psi)}{\rho\sin\psi\,\cos^{2}\psi}
    -2\tan^{2}\psi F_{2}(\rho,\psi)-2\tan\psi F_{2,\psi}(\rho,\psi)~,
    \label{U2}\\
    V_{2}&=& \frac{\rho(2\rho^{2}-\mu^{2})}{(\rho^{2}-\mu^{2})\sin^{2}\psi\,\cos\psi}
    \left[3H_{2}(\rho\psi)-\tan\psi
    H_{2,\psi}(\rho\psi)\right]
    -\frac{2\rho^{2}H_{2,\rho}(\rho\psi)}{\sin^{2}\psi\,\cos\psi}\nonumber\\
    & &
    +\frac{\rho^{3}}{\rho^{2}-\mu^{2}}F_{2,\psi}(\rho,\psi)
    -\rho^{2}\tan\psi F_{2,\rho}(\rho,\psi)
    +\frac{\pi^{4}\rho(2\rho^{2}-\mu^{2})\sin2\psi}{864\mu^{4}(\rho^{2}-\mu^{2})}\left[
    \mu^{2}(15\rho^{2}+\mu^{2})\cos6\psi\right.\nonumber\\
    & &\left. +(-8\rho^{4}+29\mu^{2}\rho^{2}+4\mu^{4})\cos4\psi
    +(184\rho^{4}+185\mu^{2}\rho^{2}+11\mu^{4})\cos2\psi
    -96\rho^{4}+99\mu^{2}\rho^{2}-16\mu^{4}\right]~.
    \label{V2}
\end{eqnarray}
\end{subequations}
The function $H_{2}$ satisfies the differential equation (\ref{Hnequation}) with
\begin{eqnarray}
    S_{2}&=&\frac{5\pi^{4}(9\rho^{2}+\mu^{2})}{48384\mu^{2}}h_{4}(\psi)
    -\frac{\pi^{4}(20\rho^{4}-71\mu^{2}\rho^{2}-23\mu^{4})}{34560\mu^{4}}h_{3}(\psi)
    +\frac{\pi^{4}(308\rho^{4}+310\mu^{2}\rho^{2}+71\mu^{4})}{24192\mu^{4}}h_{2}(\psi)\nonumber\\
    & &+\frac{\pi^{4}(432\rho^{2}+41\mu^{2})}{17280\mu^{2}}h_{1}(\psi)
    -\frac{23\pi^{4}\rho^{2}(\rho^{2}-\mu^{2})}{1728\mu^{4}}h_{0}(\psi)~,
    \label{S2}
\end{eqnarray}
where the functions $h_{n}$ are given in Eq.(\ref{hpsisin}).

The
solution of Eq.(\ref{Hnequation}) with inhomogeneity (\ref{S2}) can be written in form
 Eq.(\ref{Hnsum}) with $p_{n>4}(\rho)=0$. For each of the
non vanishing $p_{n}(\rho)$ one can add combinations of the homogeneous solution
which includes the Legendre functions as appear in Eq.(\ref{Hnsum}).
These
should be chosen such that $p_{n}(\rho)$ is regular at $\rho=1$,
and is a polynomial in $\rho$ of the lowest order possible
(in principle, the homogeneous solution is of order $2n+2$).
In fact, the particular solution for Eq.(\ref{Hnequation}) with inhomogeneity (\ref{S2})
includes  polynomials of order $4$ in $\rho$, only,
\begin{subequations}
\label{H2P}
\begin{eqnarray}
    & &H_{2}^{P}(\rho,\psi)=\pi^{4}\sum_{n=0}^{4}p_{n}(\rho)h_{n}(\psi)~,\\
    & &p_{0}(\rho=R\mu)=-\frac{23R^{4}}{13824}~,\\
    & &p_{1}(\rho=R\mu)=-\frac{432R^{2}+41}{138240}~,\\
    & &p_{2}(\rho=R\mu)=-\frac{462R^{4}+156R^{2}+71}{580608}~,\\
    & &p_{3}(\rho=R\mu)=\frac{24R^{4}-75R^{2}-23}{1658880}~,\\
    & &p_{4}(\rho=R\mu)=-\frac{9R^{2}+1}{774144}~.
\end{eqnarray}
\end{subequations}
In addition, to avoid negative powers of $\mu$,
the homogeneous part of $H_{2}$ and the gauge function $F_{2}$
should be polynomials of order $4$ in $\rho=R\mu$.
\begin{eqnarray}
    H_{2}(\rho,\psi)&\equiv&H_{2}^{P}(\rho,\psi)
    +\left[a_{0}(R^{2}-1)+b_{0}\right]\sin^{3}\psi
    +8a_{1}R^{2}(R^{2}-1)(1+\cos^{2}\psi)\sin^{3}\psi~,\label{H2}\\
    F_{2}(\rho,\psi)&\equiv& (R^{2}-1)^{2}f_{4}(\psi)
    +(R^{2}-1)f_{2}(\psi)+f_{0}(\psi)~.\label{F2}
\end{eqnarray}
We now turn to imposing boundary conditions.

\subsubsection{Conditions at the horizon}
The horizon is located at a constant $\rho=\rho_{H}$. We set
$\rho_{H}=1+\pi^{2}\epsilon/24+\epsilon^{2}\zeta_{2}$ and
solve $B(\rho_{H},\psi)=0$ for $f_{0}(\psi)$. We find
\begin{equation}
    f_{0}(\psi)=-\zeta_{2}+\pi^{4}\left[\frac{\cos8\psi}{2304}+
    \frac{\cos6\psi}{432}+\frac{\cos4\psi}{36}+\frac{7\cos2\psi}{72}+
    \frac{583}{6912}\right]~.\label{f0}
\end{equation}
The surface gravity is constant (due to the fact that the Legendre functions
of the second kind are not included in (\ref{H2})), 
\begin{equation}
    \kappa=1-\frac{5\pi^{2}\epsilon}{24}-\epsilon^{2}
    \left[6b_{0}+\frac{113\pi^{4}}{3456}\right]~.\label{kappa2}
\end{equation}

\subsubsection{Conditions at 'infinity'}
We match the near solution, (\ref{emetric2}), to the asymptotic solution,
(\ref{asymp2}), expanded to order $L^{-4}$. Comparing the expressions for  $g_{tt}$, we find that
\begin{eqnarray}
    f_{4}(\psi)&=&\frac{\pi^{4}}{8640}\left[29\cos6\psi+54\cos4\psi
    -237\cos2\psi-166\right]~,\label{2f4}\\
    f_{2}(\psi)&=&\frac{\pi^{4}}{17280}\left[15\cos8\psi+188\cos6\psi
    +1248\cos4\psi+3156\cos2\psi+2993\right]~.\label{2f2}
\end{eqnarray}
Matching the other components, as well, we find the constants
\begin{eqnarray}
    a_{0}&=&-\frac{311\pi^{4}}{13824}~,\label{2a0}\\
    a_{1}&=&\frac{89\pi^{4}}{23040}~\label{2a1}.
\end{eqnarray}
The gauge function, $W^{(2)}$, is
\begin{eqnarray}
   W^{(2)}&=&-\frac{3L^{4}\cos\psi}{32\rho^{3}}
    -\frac{\pi^{2}L^{2}}{1536\rho\sin\psi}
    (32\pi-64\psi+15\sin2\psi+7\sin4\psi+5\sin6\psi)\nonumber\\
    &&+\frac{\pi^{4}\rho}{\cos\psi}\left[-\frac{30473}{1105920}
    +\frac{6b_{0}-\zeta_{2}}{\pi^{4}}-\frac{9697\cos2\psi}{92160}
    -\frac{2581\cos4\psi}{184320}\right.\nonumber\\
        & &\left.\;\;\;+\frac{573\cos6\psi}{81920}
    +\frac{121\cos8\psi}{73728}+\frac{13\cos10\psi}{49152}
    +\frac{(\pi-2\psi)\sin2\psi}{90}\right]~.
    \label{W2Lexpand}
\end{eqnarray}

\subsubsection{Conditions at $\psi=\pi/2$}
The metric (\ref{emetric2}) should be symmetric about the plane $\psi=\pi/2$.
As a result we find that
\begin{equation}
    b_{0}=-\frac{1003\pi^{4}}{103680}+\frac{\zeta_{2}}{6}~.\label{2b0}
\end{equation}

\subsubsection{Verification of the first law}
We calculate the entropy of the black hole
\begin{equation}
    S=\frac{\pi^{2}\mu^{3}}{2G_{5}}\left[1+\frac{\pi^{2}\epsilon}{8}
    +\epsilon^{2}\left(\frac{37\pi^{4}}{960}+3\zeta_{2}\right)\right]~.
    \label{entropy2}
\end{equation}
We compare the zeroth law $T=\kappa/(2\pi)$ and
the first law $T^{-1}=\partial S/\partial M$ to find that
\begin{equation}
    \zeta_{2}=-\frac{23\pi^{4}}{1920}~.
    \label{zeta2}
\end{equation}

To summarize, the location of the horizon, the entropy, and the
temperature are
\begin{eqnarray}
    \rho_{H}&=&\mu\left[1+\frac{\pi^{2}\epsilon}{24}
    -\frac{23\pi^{4}\epsilon^{2}}{1920}\right]~.
    \label{rhoH2}\\
    S&=&\frac{\pi^{2}\mu^{3}}{2G_{5}}\left[1+\frac{\pi^{2}\epsilon}{8}
    +\frac{\pi^{4}\epsilon^{2}}{384}\right]~.
    \label{S2final}\\
    T&=&\frac{1}{2\pi\mu}\left[1-\frac{5\pi^{2}\epsilon}{24}
    +\frac{43\pi^{4}\epsilon^{2}}{1152}\right]~.
    \label{T2final}
\end{eqnarray}
The near metric is given by
\begin{subequations}
\label{near4final}
\begin{eqnarray}
    B^{N}&=&\left(1-\frac{\rho_{H}^{2}}{\rho^{2}}\right)
    \left(1-\frac{\pi^{2}\epsilon}{6}\sin^{2}\psi\left[
    \cos2\psi+3\right]
    +\frac{\pi^{4}\epsilon^{2}}{5760\mu^{2}}\left[1570\mu^{2}+17\rho^{2}
    -4(14\rho^{2}-395\mu^{2})\cos2\psi\right.\right.\nonumber\\
        & &\left.\left.+(52\rho^{2}+580\mu^{2})\cos4\psi
    -4(2\rho^{2}-25\mu^{2})\cos6\psi
    -(\rho^{2}-2\mu^{2})\cos8\psi\right]\frac{}{}\right)~,
    \label{B4final}\\
    A^{N}&=&1+\frac{\pi^{2}\epsilon}{3\mu^{2}}\left[(3\rho^{2}-\mu^{2})\cos^{4}\psi-1
    \right]+\frac{\pi^{4}\epsilon^{2}}{5760\mu^{4}}\left[1800\rho^{4}+1324\mu^{2}\rho^{2}
    -307\mu^{4}
    +4(675\rho^{4}+422\mu^{2}\rho^{2}-325\mu^{4})\cos2\psi\right.\nonumber\\
        & &\left.
    +8(135\rho^{4}+88\mu^{2}\rho^{2}-55\mu^{4})\cos4\psi
    +4(45\rho^{4}+26\mu^{2}\rho^{2}-15\mu^{4})\cos6\psi
    +5\mu^{2}(4\rho^{2}-\mu^{2})\cos8\psi\right]~,
    \label{A4final}\\
    U^{N}&=&1+\frac{\pi^{2}\epsilon}{\mu^{2}}(\rho^{2}-\mu^{2})\sin^{2}\psi\cos^{2}\psi
    -\frac{\pi^{4}\epsilon^{2}\sin^{2}\psi}{1440\mu^{4}}
    \left[-540\rho^{4}-970\mu^{2}\rho^{2}+385\mu^{4}\right.\nonumber\\
        & &\left.
    -5(144\rho^{4}+229\mu^{2}\rho^{2}-385\mu^{4})\cos2\psi
    -5(\rho^{2}-\mu^{2})(36\rho^{2}+82\mu^{2})\cos4\psi
    -55\mu^{2}(\rho^{2}-\mu^{2})\cos6\psi\right]~,
    \label{U4final}\\
    V^{N}&=& -\frac{\pi^{2}\epsilon\rho^{3}\sin\psi\cos^{3}\psi}{\mu^{2}}
    -\frac{\pi^{4}\epsilon^{2}\rho^{3}\sin2\psi}{2880\mu^{4}}
    \left[990\mu^{2}+540\rho^{2}
    +(720\rho^{2}+1207\mu^{2})\cos2\psi\right.\nonumber\\&+&\left.2(90\rho^{2}+89\mu^{2})\cos4\psi
    +25\mu^{2}\cos6\psi\right]  ~.
    \label{V4final}
\end{eqnarray}
\end{subequations}

\section{Summary and discussion}
\label{sec:summary}
In this paper we have calculated the metric of a small black hole in a five
dimensional cylinder. The metric is found  perturbatively,
with an expansion parameter  $\epsilon=\mu^{2}L^{-2}=8G_{5}M(3\pi L^{2})^{-1}$.
$M$ is the physical ADM mass and $L$ is the
circumference of the compact dimension, both  measured at infinity.
We calculate the metric, up to second order in $\epsilon$ (fourth order in the ratio $\mu/L$). 
The asymptotic solution is given in Eqs.(\ref{A1},\ref{h1},\ref{htt1final}) and in the
appendix \ref{app:asymp2}. The near solution is provided in
Eqs. (\ref{spherical},\ref{near4final}).

We obtained the following expressions for the entropy and the temperature of the black hole 
\begin{eqnarray}
    S&=&\frac{\pi^{2}L^{3}}{2G_{5}}\epsilon^{3/2}\left[1+\frac{\pi^{2}\epsilon}{8}
    +\frac{\pi^{4}\epsilon^{2}}{384}\right]~,
    \label{Sfinal}\\
    T&=&\frac{1}{2\pi\mu}\left[1-\frac{5\pi^{2}\epsilon}{24}
    +\frac{43\pi^{4}\epsilon^{2}}{1152}\right]~.
    \label{Tfinal}
\end{eqnarray}
These, and other results, presented below, were previously obtained in first order of $\epsilon$ by Gorbonos and Kol~\cite{gorbonos} and by Harmark~\cite{harmark2}.
Using the Smarr formula we find the relative binding energy
\begin{equation}
    n(\epsilon)=\frac{\pi^{2}\epsilon}{6}-\frac{\pi^{4}\epsilon^{2}}{36}~.
    \label{nfinal}
\end{equation}

We  have chosen the coordinate system such that the horizon is independent of the angle, $\psi$, between the three dimensional space and the $w$ axis. As the scale of the five dimensional radial coordinate, $\rho$, is fixed by the term of the metric, $\rho^2\sin^2\psi d\Omega^2$, the radius of the horizon is uniquely determined.  The horizon is located at $\rho=\rho_{H}$ as in Eq.(\ref{spherical}). We find that
\begin{equation}
    \rho_{H}=\mu\left[1+\frac{\pi^{2}\epsilon}{24}
    -\frac{23\pi^{4}\epsilon^{2}}{1920}\right]~.
    \label{rhoHfinal}
\end{equation}

\subsection{Discussion of the black hole - black string transition}
The question of transition from black hole to black string, as a function of the mass parameter, has been discussed at length in the literature~\cite{kol2},~\cite{harmark, nonuniform} .
One scenario proposed was the transition at the intersection of the black hole and nonuniform black string lines in the $(M, n(M))$ (mass-relative tension) phase diagram. 
We use the second order results for the small black hole to extrapolate
to such a transition.
We study three aspects of the transition; the mass-relative tension phase diagram,
 comparison of entropies, and the change of topology (using two different approaches).

A uniform black string is described by the metric
\begin{equation}
    ds^{2}_{BS}=-\left(1-\frac{2G_{5}M}{Lr}\right)dt^{2}
    +\left(1-\frac{2G_{5}M}{Lr}\right)^{-1}dr^{2}+r^{2}d\Omega_{2}^{2}+dw^{2}~.
    \label{BSmetric}
\end{equation}
The relative tension is constant, $n_{BS}=1/d-3=1/2$,
and the entropy is
\begin{equation}
    S_{BS}=\frac{1}{4G_{5}}4\pi L\left(\frac{2G_{5}M}{L}\right)^{2}
    =\frac{9\pi^{3}L^{3}}{16G_{5}}\epsilon^{2}~.
    \label{BSentropy}
\end{equation}
The entropy and relative tension of the non-uniform black string were have not been found numerically yet.\cite{nonuniform}.
Such a configuration exists for masses larger than the Gregory-Laflamme mass, which is, in terms of our variables, $\epsilon>\epsilon_{GL}$.

\begin{figure}[htb]
    \centering
    \framebox{
    \includegraphics[scale=1]{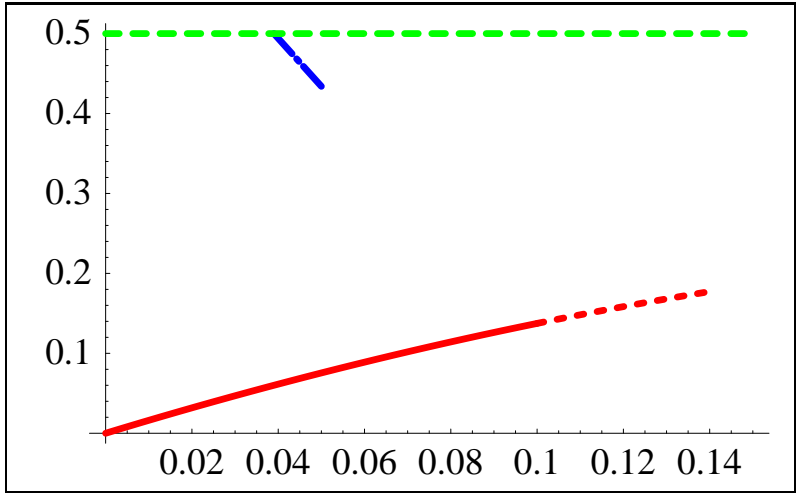}}
    \caption{\label{phasediagram}Phase diagram for the uniform black string
    (dashed green line), non-uniform black string (blue line), and black hole branches (solid red continued in dashed red in the region where the black hole turns into a nonuniform black string). }
\end{figure}

In Fig.(\ref{phasediagram}) we draw the $(\epsilon,n)$ phase diagram for the
uniform black string, the non-uniform black string (just the leading order term near the GL transition), and the black hole branches. The black hole turns into a nonuniform black string beyond the transition point, $\epsilon=0.1$. We did not include recent numerical results~\cite{kudoh2}, which, due  to numerical difficulties, have large errors~\cite{kudoh3}

\begin{figure}[htb]
    \centering
    \framebox{
    \includegraphics[scale=1]{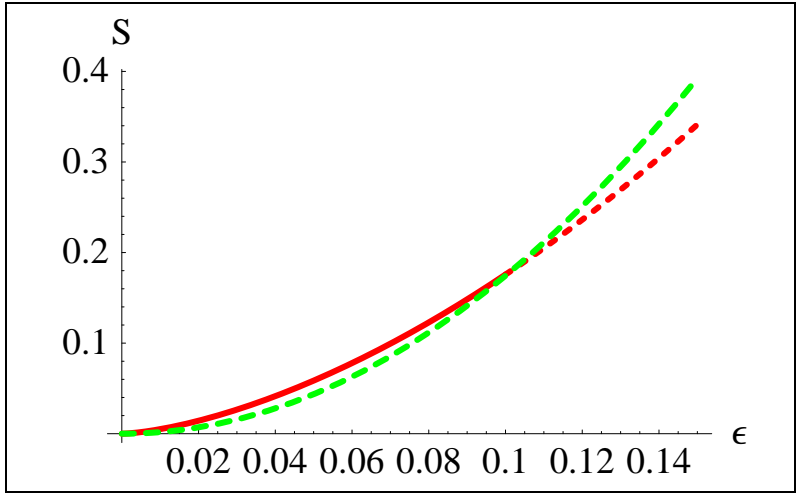}}
    \caption{\label{entropies}The entropy (in units of
    $L^{3}/G_{5}$) of the uniform black string
    (dashed green line), non-uniform black string ( dashed red line), and black hole branches (solid red line).}
\end{figure}

In Fig.(\ref{entropies}) we draw the entropies of the black objects in units of
$L^{3}/G_{5}$.
Comparing  (\ref{Sfinal}) and (\ref{BSentropy}) we find that the
enropies of the uniform black string and the black hole
are equal for $\epsilon_{BH-BS}=0.102$.
Below that value the black hole has
higher entropy, and above that value the black string has higher entropy.
This simple check suggests that the black hole will be unstable for $\epsilon>0.102$. Note again that our second order expansion in $\epsilon$ shows good convergence.  Using the first order expansion only the intersection of entropies would occur at $\epsilon=1.01$.

A transition between a black hole and a black string requires a topology change
of the horizon. This change is likely to happen when the
horizon of the black hole fills the compact dimension and touch itself.
At this point a small pinch of the horizon will turn it into a non-uniform
black string. In the notation we use, Eq.(\ref{transformation}),
this will happen when
\begin{equation}
    \rho_{H}=\frac{L}{\pi}\;\Longrightarrow \epsilon_{p}=0.0957~.
    \label{pinchpoint}
\end{equation}
The first order formula would give $\epsilon_{p}=0.094$.  Again, we see a fairly good convergence, making our result reasonably reliable.   We will see a further indication for this point in another, independent, calculation of the critical value of $\epsilon$.  

The apparent agreement of $\epsilon_p$ with  $\epsilon_{BH-BS}$ (equality of the entropies of the uniform black string and the black hole) is somewhat problematic. We see no mathematical reason for such a agreement, though it would be interesting if these points also coincided. Note that for geometric reasons the black hole branch should cease to exist above the critical value of $\epsilon$ so the stable state of the system is the uniform black string, which has a higher entropy than the nonuniform black string.  

\subsection{Discussion of the shape of the horizon}
 In terms of the coordinates we chose the horizon is located at constant $\rho$. However, this does not
mean it is spherical. Compactification breaks the $O(4)$ symmetry.  The horizon can be an oblate (elongated perpendicular
to the compact direction) or a prolate (elongated parallel to the
compact dimension) ellipsoid. There is no generally accepted method for distinguishing between the oblate and the prolate
configurations. Some authors \cite{kol,harmark} define the
'eccentricity' of the black hole as
\begin{equation}
    e=\frac{A_{||}}{A_{\perp}}-1~,\label{e}
\end{equation}
where $A_{||}$ is the maximal area of a cross section of the horizon parallel
to the compact dimension,
and $A_{\perp}$ is the maximal area of a cross section of the horizon perpendicular
to the compact dimension.
\begin{eqnarray}
    A_{||}&=\left.\int_{0}^{\pi}d\psi\int_{0}^{2\pi}d\phi\,
    \rho^{2}\sin\psi\sin\theta\sqrt{U(\rho,\psi)}\right|_{\rho=\rho_{H},\theta=\pi/2}
    &=2\pi\rho_{H}^{2}\int_{0}^{\pi}d\psi\,\sin\psi\sqrt{U(\rho_{H},\psi)}
    ~,\label{A||}\\
    A_{\perp}&=\left.\int_{0}^{\pi}d\theta\int_{0}^{2\pi}d\phi\,
    \rho^{2}\sin^{2}\psi\sin\theta\right|_{\rho=\rho_{H},\psi=\pi/2}&=4\pi\rho_{H}^{2}
    ~.\label{Aperp}
\end{eqnarray}
A prolate (oblate) horizon has positive (negative) eccentricity.
We use the near solution (\ref{emetric2}) to find that
$e=4\pi^{4}\epsilon^{2}/135$. We see that the horizon is prolate.

Another possible measure of the eccentricity is given by the intrinsic Ricci scalar of the
horizon, which is calculated in terms of the three metric
\begin{equation}
    ds^{2}_{H}=\rho_{H}^{2}\left[U(\rho_{H},\psi)d\psi^{2}+
    \sin^{2}\psi\left(d\theta^{2}+\sin^{2}\theta d\phi^{2}\right)\right]
    ~.\label{horizonmetric}
\end{equation}
Using the near solution, the intrinsic curvature of
(\ref{horizonmetric}) is
\begin{equation}
    R^{(3)}=\frac{6}{\rho_{H}^{2}}\left[1+\frac{2\pi^{4}\epsilon^{4}}{135}
    (1+5\cos2\psi)\right]~.\label{intrinsicR}
\end{equation}
It is maximal at $\psi=0$, which indicates that the horizon is indeed prolate.

One might speculate that a prolate horizon will tend to grow in
the periodic direction and turn into a nonuniform black string. However, the circumference of the compact direction
grows as well~\cite{kol} and the transition into a black string is not
obvious. In \cite{kol} it is suggested to measure the 'inter polar
distance', which is the proper distance between the poles of the
prolate horizon along the compact dimension. In principle, to calculate this distance we need to break the integral defining this distance into two parts.  In the first part we should use the near solution and in the
other we should use the asymptotic solution. In the near horizon region one
should integrate along $\psi=0$ from $\rho=\rho_{H}$ to some
arbitrary point $\rho=z$. In the asymptotic region we should integrate
along $r=0$ from $w=w(z)$ to $w=L/2$
\begin{equation}
    \frac{1}{2}L_{poles}=\int_{\rho_{H}}^{z}\sqrt{g_{\rho\rho}^{N}}d\rho
    +\int_{\frac{L}{\pi}\arcsin \pi z/L}^{L/2}\sqrt{g_{ww}^{A}}dw
    ~.\label{Lpoles}
\end{equation}
In \cite{kol} it was found that the zeroth order approximation is $L_{poles}=L$,
which means that the circumference grows enough to make room for the black hole. 

 If the mass ($\epsilon$) value, at which the black hole fills the compact dimension, is small enough then it is possible that the integral over the near solution is sufficient in (\ref{Lpoles}), or, in other words, we can choose $z=L$.  Then we obtain
 \begin{equation}
    \frac{L_{\rm poles}}{L}=\frac{149}{60\pi}-\frac{17 \pi}{24}\epsilon -\frac{1003\pi^3}{1440}\epsilon^2+O(\epsilon^3)
    ~.\label{Lpoles2}
\end{equation}
First of all, we see that the distance between the poles of the horizon is a decreasing function of  $\epsilon$. If we solve the equation $L_{\rm poles}=0$ we obtain $\epsilon_{cc}=0.146$, which is in a rough agreement with values we obtained for the critical value, $\epsilon_c$, above.  Taking the first order term only we would obtain $\epsilon_{cc}=0.355$. Clearly, the convergence of $\epsilon_{cc}$ is not as good as those of $\epsilon_c$ and $\epsilon_p$.  One reason for the poorer convergence is that for very small $\epsilon$ (\ref{Lpoles2}) must fail, as a significant contribution coming from the asymptotic solution was omitted. It should work, however in the region where $L_{\rm poles}\ll L$. Fortunately, in this region the $\epsilon$-expansion is still fairly reliable and the rough agreement of $\epsilon_c\simeq\epsilon_{cc}$ is encouraging.

\begin{acknowledgments}
This work is supported in part by the U.S. Department of Energy Grant
No. DE-FG02-84ER40153.
We thank Richard Gass and Cenalo Vaz for fruitful discussions. We also thank Barak Kol, Hideaki Kudoh, Evgeny Sorkin, and Toby Wiseman for their comments, which helped us improving this work.
\end{acknowledgments}

\appendix
\section{Second order asymptotic solution}
\label{app:asymp2}
The second order asymptotic solution is given by Eqs.(\ref{h2})
\begin{subequations}
\label{h2A}
\begin{eqnarray}
    h_{tt}^{(2)}&=&h_{tt}^{(H)}+\bar{h}^{(2)}_{tt}~,\\
    h_{rr}^{(2)}&=&-\frac{r}{2}h_{tt,r}^{(H)}+\bar{h}^{(2)}_{rr}~,\\
    h_{ww}^{(2)}&=&-\frac{r}{2}h_{tt,r}^{(2)}+2W_{,w}^{(2)}+\bar{h}_{ww}^{(2)}~,\\
    h_{rw}^{(2)}&=&W_{,r}^{(2)}+\bar{h}_{rw}^{(2)}~.
\end{eqnarray}
\end{subequations}
The the components of the inhomogeneity, $\bar{h}^{(2)}$, are
\begin{subequations}
\label{h2ANH}
\begin{eqnarray}
    \bar{h}_{tt}^{(2)}&=&-\frac{1}{4}h^{2}+W\,h_{,w}
    ~,\\
    \bar{h}_{rr}^{(2)}&=&\frac{r^{2}}{16}\left[(h_{,r})^{2}
    +(h_{,w})^{2}\right]-\frac{r}{8}\left[h\,h_{,r}
    +4W\,h_{,rw}\right]+(W_{,r})^{2}-r\,K_{,w}
    ~,\\
    \bar{h}_{ww}^{(2)}&=&\frac{r^{2}}{16}\left[(h_{,r})^{2}
    +(h_{,w})^{2}\right]-\frac{r}{2}\left[2W_{,w}h_{,r}
    +W\,h_{,rw}\right]+\frac{3}{16}h^{2}+(W_{,w})^{2}-r\,K_{,w}
    ~,\\
    \bar{h}_{rw}^{(2)}&=&K-\frac{r}{16}\left[h\,h_{,w}+8W_{,r}h_{,r}\right]
    +W_{,w}W_{,r}~,
\end{eqnarray}
\end{subequations}
where $h=h_{tt}^{(1)}$, $W=W^{(1)}$, and $K_{,ww}=h\,h_{,rw}/8-h_{,w}h_{,r}/4$,
\begin{eqnarray}
    h&=&h_{tt}^{(1)}=\frac{\pi L \sinh\frac{2\pi r}{L}}
    {r(\cosh\frac{2\pi r}{L}-\cos\frac{2\pi w}{L})}~,\\
    W&=&W^{(1)}=-\frac{\pi L\sin\frac{2\pi w}{L}}
    {12(\cosh\frac{2\pi r}{L}-\cos\frac{2\pi w}{L})}
    \left[\frac{3L}{2\pi r}\sinh\frac{2\pi r}{L}-2\cos\frac{2\pi w}{L}+5\right]~,
    \label{W1final}\\
     K&=&\frac{\pi L^{3}}{16r^{3}}\coth\frac{2\pi r}{L}
    \left[1-\frac{2\pi r}{L}\coth\frac{2\pi r}{L}\right]
    \arctan\left(\tan\frac{\pi w}{L}\coth\frac{\pi r}{L}\right)
    +\frac{\pi L^{3}\sin\frac{2\pi w}{L}
    \left[1-\frac{2\pi r}{L}\coth\frac{2\pi r}{L}\right]}
    {32r^{3}\left[\cosh\frac{2\pi r}{L}-\cos\frac{2\pi w}{L}\right]}~.
\end{eqnarray}
The function $W^{(1)}$ was chosen such that $W(r\gg L,w)={\cal
O}(r^{1})$. It satisfies Eqs. (\ref{W0},\ref{W2}), and it has a
structure similar to $h_{tt}^{(1)}$.

$K$ is an odd function of $w$, but $K(r,w=L/2)\neq0$. However,
$Z_{2}$-symmetry implies that $h_{rw}^{(2)}(r,w=L/2)=0$. This
means that the gauge function, $W^{(2)}$, has nontrivial boundary
conditions at $w=\pm L/2$
\begin{equation}
    W_{,r}^{(2)}(r,w=L/2)=-W_{,r}^{(2)}(r,w=-L/2)=
    -\frac{\pi^{2} L^{3}}{32r^{3}}\coth\frac{2\pi r}{L}
    \left[1-\frac{2\pi r}{L}\coth\frac{2\pi r}{L}\right]~.
    \label{W2L2}
\end{equation}

To match the asymptotic and the near solution we have to
transform the asymptotic solution to the $(\rho,\psi)$ coordinates, using
(\ref{transformation}), and expand to order $L^{-4}$.
We also expand the gauge function $W^{(2)}$ as
\begin{equation}
    W^{(2)}(\rho,\psi)=L^{4}\left[W^{(2)}_{0}
    +\frac{\pi^{2}}{L^{2}}W^{(2)}_{2}+\frac{\pi^{4}}{L^{4}}W^{(2)}_{4}\right]~.
    \label{W2near}
\end{equation}
The asymptotic metric in the $(\rho,\psi)$ coordinates, expanded to order $L^{-4}$ is
\begin{subequations}
\label{asymp2}
\begin{eqnarray}
    B^{A}&=&1-\frac{\mu^{2}}{\rho^{2}}-\frac{\pi^{2}\mu^{2}\sin^{2}\psi
    }{6L^{2}}(\cos2\psi+3)+\frac{\pi^{4}\mu^{2}\rho^{2}\sin^{4}\psi}{45L^{4}}
    (\sin^{4}\psi-6\cos^{4}\psi)\nonumber\\
    & &-\frac{\pi^{2}\mu^{4}}{24L^{2}\rho^{2}}(\cos4\psi+4\cos2\psi-3)\nonumber\\&
    +&\frac{\pi^{4}\mu^{4}}{5760L^{4}}(15\cos8\psi+100\cos6\psi+528\cos4\psi
    +1636\cos2\psi+1553)~,
    \label{Basymp2}\\
    A^{A}&=&1+\frac{\pi^{2}\rho^{2}\cos^{4}\psi}{L^{2}}
    +\frac{\pi^{4}\rho^{4}\cos^{6}\psi}{L^{4}}
    -\frac{\pi^{2}\mu^{2}}{24L^{2}}(\cos4\psi+4\cos2\psi+11)\nonumber\\
   & +&\frac{\pi^{4}\mu^{2}\rho^{2}}{1440L^{4}}(5\cos8\psi
    +26\cos6\psi+176\cos4\psi+422\cos2\psi+331)\nonumber\\
 &   +&\frac{\mu^{4}\cos\psi}{16\rho^{4}}(32\rho^{4}W^{(2)}_{0,r}-9\cos\psi)
    ~,
    \label{Aasymp2}\\
    U^{A}&=&1+\frac{\pi^{2}}{L^{2}}\left[\rho^{2}\cos^{2}\psi\sin^{2}\psi
    -\frac{\mu^{2}\sin\psi\cos^{2}\psi(5\sin3\psi-37\sin\psi)}{12}
    -2\mu^{2}\frac{\sin\psi}{\rho} W_{2,\psi}(\rho,\psi)\right]~,
    \label{Uasymp2}\\
    V^{A}&=& -\frac{\pi^{2}}{L^{2}}\left[\rho^{3}\sin\psi\cos^{3}\psi
    +\mu^{2}\rho\sin\psi\cos^{3}\psi(\sin^{2}\psi+2)
    +\mu^{2}\rho\sin\psi W_{2,\rho}(\rho,\psi)
    -\mu^{2}\cos\psi W_{2,\psi}(\rho,\psi)\right] ~.
    \label{Vasymp2}
\end{eqnarray}
\end{subequations}

\end{document}